\documentclass[review,12pt,authoryear]{elsarticle}

\usepackage{epsfig}

\usepackage{amssymb}

\usepackage{amsbsy}
\usepackage{color}

\usepackage{eurosym}

\usepackage{rotating}
\journal{European Journal of Operational Research}

\begin{document}

\begin{frontmatter}

\title{Default Probability Estimation via Pair Copula Constructions}

\author[label1]{Luciana Dalla Valle }
    \author[label2]{Maria Elena De Giuli}
    \author[label2]{Claudia Tarantola}
 \author[label3]{Claudio Manelli}

 \address[label1]{School of
Computing and Mathematics, Plymouth University, UK}
 \address[label2]{Department of Economics and Management, University of Pavia, Italy}
\address[label3]{List S.p.A., Piazza Duomo 57, 27058 Voghera, Italy}

\cortext[cor1]{Address for correspondence: Claudia Tarantola,
Department of Economics and Management, University of Pavia, Via
San Felice 5, 27100, Pavia, Italy,
Tel:+39-0382-986213, Fax:+39-0382-304226 \\
E-mail:~claudia.tarantola@unipv.it}

\address{}

\begin{abstract}
In this paper we present a novel
approach for firm default probability estimation. The
methodology is   based on multivariate contingent claim analysis
and pair copula constructions. For each considered firm, balance
sheet data are used to assess the asset value, and to compute  its
default probability. The asset pricing function is expressed via a
pair copula construction, and it is approximated via Monte Carlo
simulations. The methodology is illustrated through an application
to the analysis of both
 operative and
defaulted firms.

\end{abstract}


\end{frontmatter}

\section{Introduction}
%Default risk is defined as the risk of loss when in a financial
%contract a debtor (in our case a firm) does not fulfil its
%commitments and a default event takes place.
Default risk is defined as the
risk of a loss when  a debtor (in our case a firm) does not fulfil
its commitments in a financial contract, and a default event takes
place.
The probability of default (PD) is the probability that a
default happens.

Following the growing financial uncertainty, there has been
intensive research by  institutions, regulators and academics to
develop models for firm evaluation and PD estimation.
 The existing methodologies differ on the
available information and data used for assessing the firm value.
They can be broadly classified in models based on market data and
on accounting data.

Within the market data based models, the most popular are
structural models; see \cite{Merton70, Merton74, Merton77}
 and their extensions; for recent and complete
reviews, see e.g. \cite{Ji10}, \cite{Laajimi12}, or
\cite{Sundaresan13}. The asset value is considered to be exogenous
and it is treated as the underlying asset in a contingent claim
framework.
  A common assumption is that  the asset value  follows a geometric Brownian
  motion, and its drift and volatility coefficients do not depend on the capital structure of the firm.
Black  and Scholes'
 formula is applied to compute the asset price,
and consequently the PD can  be easily estimated, see
\cite{Black1973}.

The second class of models  use   accounting data and financial
ratios to evaluate the firm value, and its PD. They origin from
the   works of \cite{Beaver68} and \cite{Altman68} who developed
univariate and multivariate models, based on linear discriminant
analysis, to predict the default of specific firms by using a set
of financial ratios. Another commonly used default prediction
model
 is based on  logistic regression, as proposed by \cite{Ohlson80}.

The previous models have been analysed both in  a classical and in
a Bayesian framework. For some recent works in the classical
framework, see e.g. \cite{BharShum08}, \cite{DGFanMa08},
\cite{Kreinin08},
\cite{SuHuang10}, \cite{AltFarKal11},
\cite{BoTaWaYa11,BoLiWaYa13}, \cite{BhiGulRL14},
\cite{Leow and Crook14},
\cite{Tobback et al14},
  and references therein. For the Bayesian analysis
 see e.g. \cite{Kiefer09, Kiefer10, Kiefer11}, \cite{PaSiKim10}, \cite{Tasche11}, \cite{KaMo12}, \cite{Orth13}, \cite{Liu et al.15}, and references therein.

A  popular and  efficient tool in risk management is  the copula
function, introduced by \cite{Sklar59}. The advantage of copulas
is the ability to obtain the joint multivariate distribution
embedding the variable's dependence structure.
 Unfortunately, while
there is a wide range of possible alternative copula functions for
the bivariate case, in the multivariate setting the use of
families different from Normal and Student's t is rather scarce,
due to computational and theoretical limitations. For this reason
\cite{Joe96} introduced Pair Copula Constructions (PCCs) to
represent complex structures of dependence among multivariate
data. PCCs constitute a flexible and very appealing tool for
financial analysis, see e.g. \cite{VazMenCam10}, \cite{MinCza10},
\cite{AlAsMAPoSi13}, \cite{DiBrCzKu13}, \cite{BerCz13}, and
reference therein. A collection of potentially different bivariate
copulas is used to construct the joint distribution of interest
via PCCs, allowing to represent different types and strengths of
dependence in an easy way.

In this paper we propose a novel approach for PD estimation, that
combines features of both
 structural and accounting based models.
 We consider a contingent
claim model based on balance sheet data, where the dynamic of the
equity is described via a  PCC and calculated using  Monte Carlo
simulations.
We apply Bayesian
parametric mixture models in the new context of vine marginal
modelling, for balance sheet data of defaulted and non-defaulted
firms.
The
 PD is obtained in  a fairly straightforward way from the equity distribution.

The outline of the paper is the following. In Section 2 we briefly
present
 copula models and PCCs.
 In Section 3 we  introduce
 a novel balance sheet multivariate contingent claim model for PD estimation based on PCCs.
 In Section 4 the
 model estimation methodology is presented.
 Section 5 describes the application of the proposed methodology to
 the PD estimation of  defaulted and  operative companies.
 Finally, concluding remarks are given in Section 6.

\section{Background and Preliminaries}\label{pcc_parag}

\subsection{Copula Function}

Copulas are very popular and appealing statistical tools, that
allow us to describe complex multivariate  patterns of dependence
  binding together the  marginal  distributions.
They are
  applicable to a wide variety of fields, such as economics, finance  and marketing;
for a  review see e.g. \cite{Jaworski10}.

A copula is a multivariate distribution function with uniform marginals
 on the interval $[0,1]$.
 Once applied to the univariate marginals, it  returns
 the multivariate joint distribution, en\-clo\-sing all the information about the dependence structure of the variables.
Thus, the use of copulas allows us to split the distribution of a
random vector into its individual marginal components, and
 the dependence structure is modelled through the copula function without losing information;
  for more details see e.g. \cite{Joe97} and \cite{Nelsen99}.

Sklar's theorem is  the most important result in copula theory. It
states that, given  a vector of random variables $\textbf{X} =
(X_1, \ldots, X_d)$,
 with $d$-di\-men\-sio\-nal joint cumulative distribution function $F(x_1, \ldots, x_d)$ and marginal
 cumulative distributions $F_m(x_m)$ with $m=1, \ldots, d$,  there exist a $d$-di\-men\-sio\-nal copula $C$ such that
\begin{equation}
F(x_1, \ldots, x_d) = C(F_1(x_1), \ldots, F_d(x_d); \boldsymbol{\theta}),
\label{copula}
\end{equation}
where $\boldsymbol{\theta}$ denotes the set of parameters of the
copula. To simplify the notation, in the remainder of the paper,
we  set $$C(F_1(x_1), \ldots, F_d(x_d))=C(F_1(x_1), \ldots,
F_d(x_d); \boldsymbol{\theta}).$$

For an absolutely continuous joint distribution $F$ with strictly
increasing continuous marginal distribution functions, the $d$-dimensional copula is uniquely defined.
Conversely, according to Nelsen's corollary, the inversion method
allows us to express the copula in the following way
$$
C(u_1, \ldots, u_d)=F(F_1^{-1}(u_1), \ldots, F_d^{-1}(u_d)),
$$
where $F_1^{-1}, \ldots, F_d^{-1}$ are the generalised inverse
functions of the marginals.

The joint density function is
$$
f(x_1, \ldots, x_d) = c(F_1(x_1), \ldots, F_d(x_d)) \cdot f_1(x_1)
\cdots f_d(x_d),
$$
where $c(F_1(x_1), \ldots, F_d(x_d))$ is the $d$-variate copula
density, provided its existence.

In this paper, we fit the data
into a given model following a parametric approach. Nonparametric methods for copula density estimation also exist, see
e.g. \cite{Sancetta and Satchell}, \cite{Shen et al} and
\cite{Kauermann et al}.

The existing literature on copulas mainly focuses on the bivariate
case. In the multivariate case,
  Normal and Student's t copula are the most popular, while the use of other multidimensional copulas is rather limited, due to the complexity
 of their construction, see e.g. \cite{AasBerg09}.
 However, Normal and Student's t copula are often not flexible enough to represent the dependence
  structure  of the data.
 Hence, multivariate extensions of Archimedean copulas were proposed in
   the form of partially nested Archimedean copulas by \cite{Joe97} and \cite{Whelan04}; hierarchical Archimedean copulas by \cite{Savutrede06}; and multiplicative
    Archi\-me\-dean copulas by \cite{Morillas05} and \cite{Liebscher06}. Nevertheless, these multivariate extensions
   imply additional restrictions on the parameters that limit their
   flexibility.   A possible solution to this problem is provided by
   PCCs, that will be described in the following section.

\subsection{Pair Copula Constructions \label{Dvine}}

We now briefly introduce PCCs, the related notation and
terminology; for more details see e.g. \cite{Czado10}. PCCs were
 originally proposed by \cite{Joe96}, and later discussed in
detail by \cite{BedCoo01, BedCoo02}, \cite{KurCo06} and
\cite{AasCzFrBa09}. For some
recent works in a parametric and nonparametric framework see e.g.
\cite{MinCza10}, \cite{Bauer et al}, \cite{NiJoeLi12},
 \cite{Weiss and Scheffer},
 \cite{Haff and Segers}.

 A PCC represents the
 complex pattern of dependence of multivariate data via a cascade of bivariate
 copulas, and
 permits to construct flexible high-dimensional copulas by using only bivariate copulas
  as building blocks, see \cite{AasCzFrBa09}.
Therefore, the joint distribution is obtained on the basis of
bivariate pair copulas, that may be conditional on a specific set
of variables, allowing to model the dependence among the
marginals.

In order to obtain a PCC we proceed as follows. First of all we
factorise the joint distribution $f(x_1, \ldots, x_d)$ of the
random vector $\textbf{X} = (X_1, \ldots, X_d)$   as a product
of conditional  densities
\begin{equation}\label{cond_fact}
    f(x_1, \ldots, x_d) = f_d(x_d) \times f_{d-1|d}(x_{d-1}|x_d) \times \ldots \times f_{1|2 \cdots d}(x_1|x_2, \ldots, x_d).
\end{equation}
The factorisation in (\ref{cond_fact}) is unique up to re-labeling
of the variables, and it
 can be reexpressed in terms  of a product of bivariate
copulas. By   Sklar's theorem the joint distribution of the
subvector $(X_d,X_{d-1})$ can be expressed in terms of a copula
density
\begin{eqnarray*}\label{biv_joint} f(x_{d-1}, x_d) = c_{d-1, d}
(F_{d-1}(x_{d-1}), F_d(x_d)) \times f_{d-1}(x_{d-1}) \times f_{d}
(x_{d}), \end{eqnarray*} where $c_{d-1, d}(\cdot, \cdot)$ is an
arbitrary bivariate copula (pair copula) density. Hence, the
conditional density of  $X_{d-1}|X_d$  can be easily rewritten as
\begin{equation}\label{cond_d-1_d} f_{d-1|d}(x_{d-1}|x_d) =
c_{d-1, d} (F_{d-1}(x_{d-1}), F_d(x_d)) \times f_{d-1}(x_{d-1}).
\end{equation}
 Through a straightforward  generalisation of
equation (\ref{cond_d-1_d}), each term in (\ref{cond_fact}) can be
decomposed into the appropriate pair copula times a conditional
marginal density.
More precisely, for  a generic element $X_\jmath$ of the vector
$\textbf{X}$ we obtain
 \begin{equation}\label{cond_cop}
    f_{x_{\jmath}|\textbf{v}}(x_\jmath|\textbf{v}) = c_{x_\jmath, v_\ell|\textbf{v}_{-\ell}}(F_{x_\jmath|\textbf{v}_{-\ell}}
    (x_\jmath|\textbf{v}_{-\ell}), F_{v_\ell|\textbf{v}_{-\ell}}(v_\ell|\textbf{v}_{-\ell}))
    \times f_{x_\jmath|\textbf{v}_{-\ell}}(x_\jmath|\textbf{v}_{-\ell}),
\end{equation}
where $\textbf{v}$ is the conditioning vector,
 $v_\ell$ is
 a generic component  of $\textbf{v}$, $\textbf{v}_{-\ell}$ is  the vector
  $\textbf{v}$ without the component $v_\ell$, $F_{x_{\jmath}|\textbf{v}_{-\ell}}(\cdot| \cdot)$ is the conditional distribution of
  $x_{\jmath}$ given $\textbf{v}_{-\ell}$,
  and $c_{x_{\jmath}, v_\ell|\textbf{v}_{-\ell}}(\cdot, \cdot)$ is the conditional pair copula density.
\noindent The $d$-dimensional joint multivariate
  distribution function can hence be expressed as a product of bivariate copulas and marginal
   distributions  by recursively plugging equation (\ref{cond_cop}) in equation  (\ref{cond_fact}).
Such decomposition is named PCC, as introduced by \cite{Joe96}.

Note that the PCC decomposition in
(\ref{cond_cop}) is based on the simplifying assumption that the
conditional copulas  depend on the conditioning variables only
indirectly through the  conditional distribution functions
 that constitute their arguments. However, as demonstrated by \cite{HaffAasFri10},
  the simplified PCC is a good approximation, even when the simplifying assumption is far from being fulfilled by the actual model.

The PCC is order dependent. A different choice of the variable
order leads to a different PCC and to a different factorisation of
the joint multivariate distribution.

Furthermore, given a specific
   factorisation there are still many different parameterisations.
For high-dimensional distributions, the  number of possible PCCs
is very high, see \cite{Czado10} and \cite{Morales11}. Hence a
suitable representation of all of them is necessary. For this
reason, \cite{BedCoo01, BedCoo02} introduced  Regular vines
(R-vines) as a pictorial representation of PCCs.
%
%Regular vines are a particular type of graphical models, that uses
%  a nested set of trees  to represent the decomposition of the joint
%distribution into its bivariate components, incorporating the
%dependence structure of the variables of interest. The class of
%regular vines is still very general and embraces a large number of
%possible pair-copula decompositions. Here we consider  a special
%case of regular vines named D-vines; for more details  see
%\cite{KurCo06}. D-vines are to be preferred over the more general
%class of R-vines because of their simple analytical formulation
%and, unlike C-vines, they do not assume the existence of a
%particular node dominating the dependencies.
%

R-vines are a particular type of
graphical models, that use
  a nested set of trees  to represent the decomposition of the joint
distribution into its bivariate components, incorporating the
dependence structure of the variables of interest. %The class of
%regular vines is still very general and embraces a large number of
%possible pair-copula decompositions.
 Two special cases of
R-vines are Canonical vines (C-vines) and Drawable vines
(D-vines), see \cite{KurCo06}. Here we consider a four dimensional
problem, for which R-vines are either C-vines or D-vines.
 We concentrate on D-vines because, differently from C-vine, they do not assume the existence of a
particular node dominating the dependencies.

 A vine ${\cal V}(d)$ on $d$ variables is a
nested set of  trees  $T_1, \ldots,
T_{d-1}$.
The edges of tree $T_\tau$ are the nodes of tree
$T_{\tau+1}$, $\tau = 1, \ldots, d-1$.
%, and  each tree has the maximum number of edges.
In a  R-vine,  if two edges of tree $T_\tau$ share a common node,
they are represented in tree $T_{\tau+1}$ by nodes joined by an
edge. A D-vine is a R-vine where all nodes do not have degree
higher than $2$, that is each node is connected to no more than
two other nodes.

\begin{figure}[htbp]
\begin{center}
 \includegraphics[width=10cm]{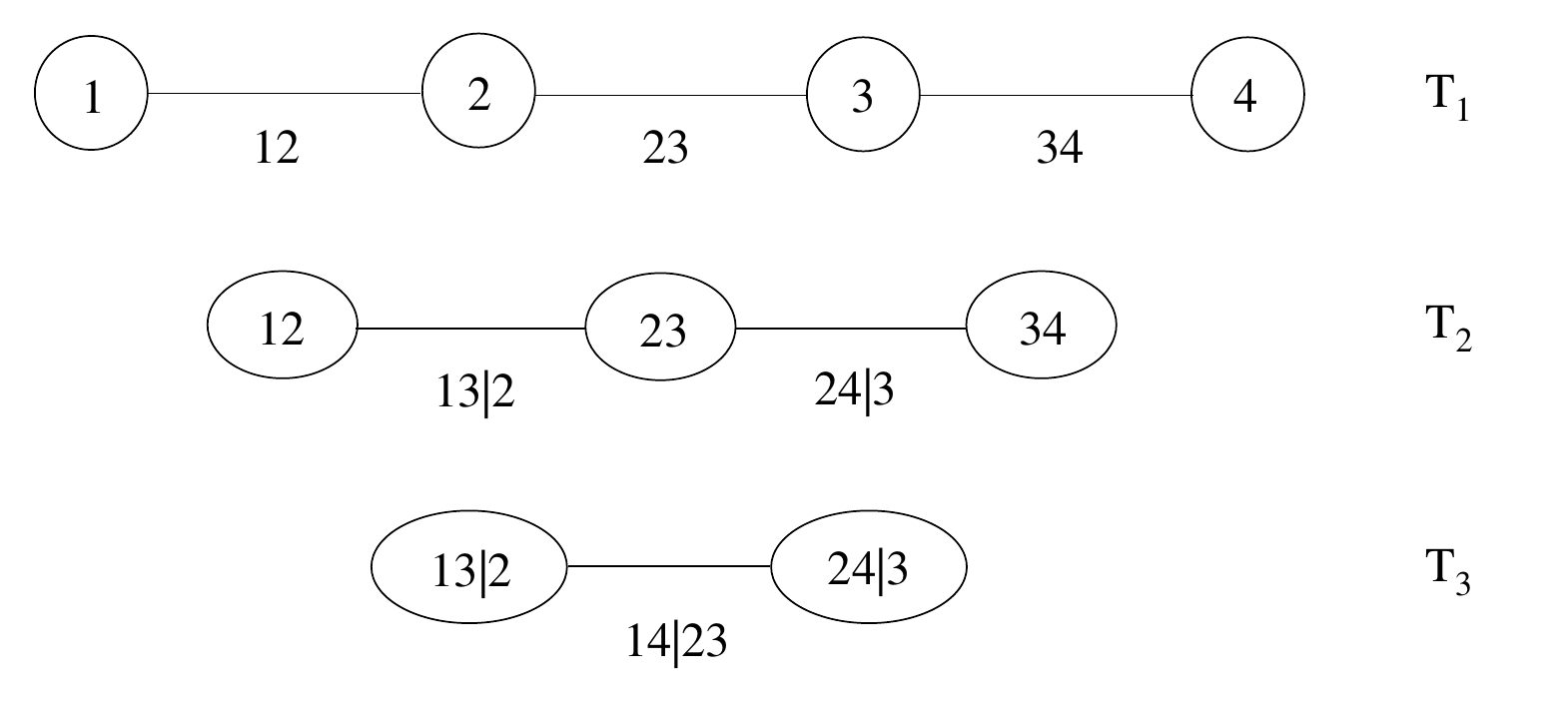}
 \caption{4-dimensional  D-vine graphical representation}\label{D_vine}
\end{center}
\end{figure}

In a D-vine, each node corresponds to a variable or a set of variables.
A  pair-copula
density is associated to any  edge, with the  edge label indicating the subscript of the
pair-copula density.
An example of a $4$-dimensional D-vine is provided by
  Figure \ref{D_vine}.
The first tree is constructed ordering the variables according to
their pairwise dependence, where the first two nodes correspond to
the variables with the strongest association, and so on; the
dependencies between nodes $\{1\}$ and $\{2\}$, between $\{2\}$
and $\{3\}$, and between $\{3\}$ and $\{4\}$ are modelled using
bivariate copula distributions. In the second tree, the
conditional dependencies between nodes $\{1,2\}$ and $\{2,3\}$,
and between $\{2,3\}$ and $\{3,4\}$ are modelled via pair copula
densities. In the third tree, the conditional dependence between
nodes $\{1,3|2\}$ and $\{2,4|3\}$ is modelled via a pair copula
density.

Using the D-vine representation, the joint density can be
decomposed in terms of conditional copula densities (identified by
the labels of the edges in the considered trees) times the
marginal densities of the examined variables. The joint density
for the D-vine represented in Figure \ref{D_vine}   is given by
$$
f(x_1, \ldots, x_4) = \prod_{\tau=1}^4 f_\tau (x_\tau) \times
c_{12}  \times c_{23}  \times c_{34}  \times c_{13|2} \times
c_{24|3} \times c_{14|23},
$$
where $c_{ab}=c_{ab}(F(x_a),F(x_b))$.

More generally, the density of a D-vine of dimension $d$ takes the
form
\begin{eqnarray*}
f(x_1, \ldots, x_d) &=& \prod_{\tau=1}^d f_\tau (x_\tau) \times
\\
     & & \prod_{j=1}^{d-1} \prod_{i=1}^{d-j}
 c_{i,i+j|i+1, \ldots, i+j-1}(F(x_i|x_{i+1}, \ldots, x_{i+j-1})
      , F(x_{i+j}|x_{i+1}, \ldots, x_{i+j-1}))
\end{eqnarray*}
which is the product of $d$ marginal densities $f_\tau$ and
$d(d-1)/2$ bivariate copulas \linebreak $c_{i,i+j|i+1, \ldots,
i+j-1}(\cdot, \cdot)$ evaluated at the conditional distribution
functions $F(\cdot|\cdot)$.

\begin{figure}[htbp]
\begin{center}
 \includegraphics[width=10cm]{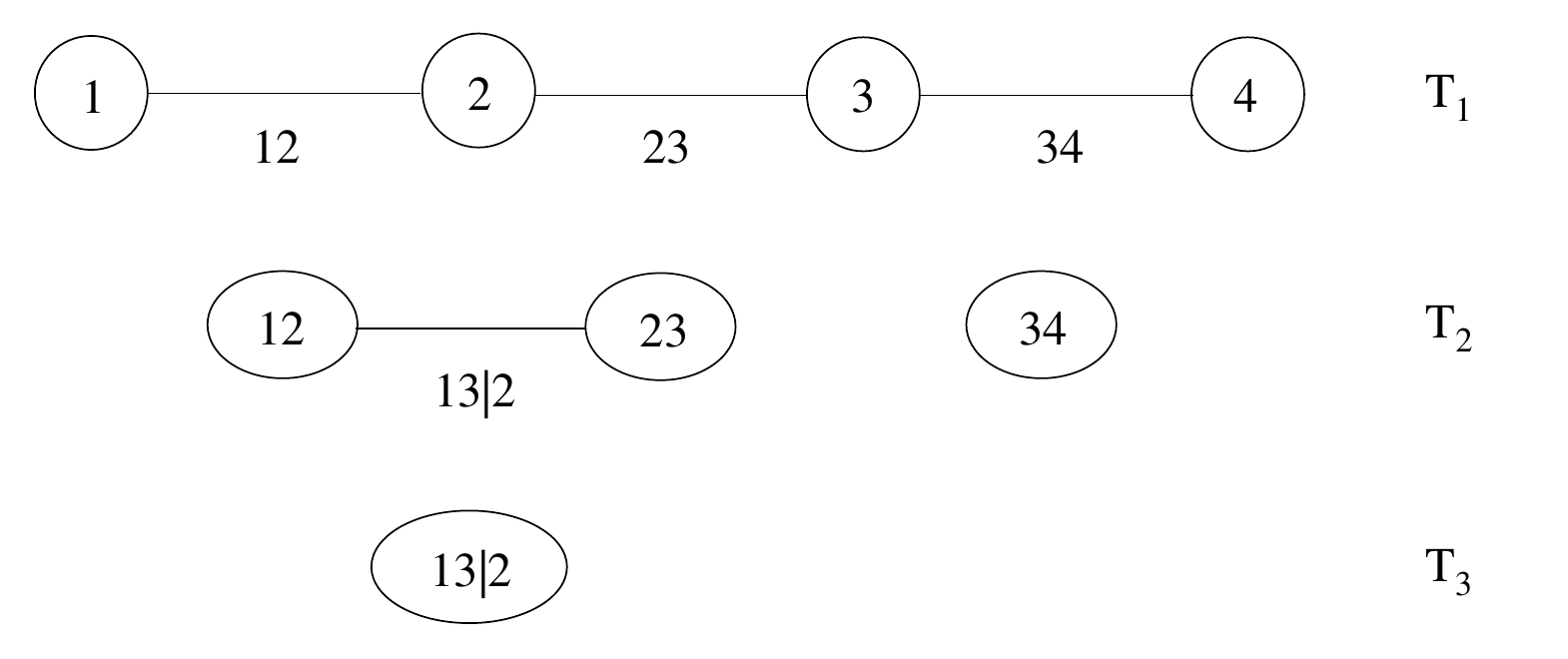}\\
  \caption{Forest vine}\label{D_vine_simpl}
\end{center}
\end{figure}

If marginal or conditional independence between pairs of variables holds, the corresponding pair copulas are equal
 to one and hence the PCC and joint density simplify accordingly.
The case of independence is depicted in the corresponding vine by
missing edges between nodes, obtaining a forest vine, as shown in
Figure \ref{D_vine_simpl}. Here, conditional independence between
variables  $2$ and $ 4 $  given $3$
 is represented by the missing edge $\{ 2, 4 | 3 \}$, which reduces the number of levels of the PCC.

\section{A Balance Sheet Multivariate Contingent Claim Model  \label{the model}}

In this paper we propose a novel
contingent claim model for PD estimation via  PCCs on  balance
sheet data, that refines and improves Merton's analysis. Our
approach allows us to evaluate, at any time $t$, the company
ability to service its debts, and consequently to efficiently
predict its PD in a flexible way.
 In the following Section we
describe the main characteristics of Merton's model, and
introduce our  approach.

\subsection{Merton's Model}

According to \cite{Merton70, Merton74, Merton77}, the evaluation
of the firm  total assets  $A_t$ is based on the structural
variables  equity $E_t$ and bond $B_t$,
$$
A_t=E_t+B_t.
$$

 \noindent A very common assumption  is that the
value $A_{t}$ of the firm follows a geometric Brownian motion
$$
\mathrm{d}A_{t}=\mu_A A_{t}\,\mathrm{d}t+\sigma_A
A_{t}\,\mathrm{d}W_{t},
\label{A(t) dyn1}%
$$
where   $\mu_A$ is the instantaneous  expected return of the
asset, $\sigma_A$ is the volatility,  and $W_{t}$ is a Wiener
process. Under the assumptions of market efficiency,
 no arbitrage opportunity and continuous
hedging, the  market value of equity satisfies
\begin{equation}
 E_t = A_t N(d_1)-e^{-rT} D N(d_2),
\label{merton1}%
\end{equation}
where  $r$ is the risk free interest rate, $N(\cdot)$ is the
cumulative standard Normal distribution function, $D$ is the face
value of bond at maturity $T$, $d_1$ is given by
$$
 d_1=\frac{\log(A_t/D)+(\mu_A+0.5\sigma^2_A) T }{\sigma_A\sqrt{T}}
\label{merton_d1}%
$$
and $d_2=d_1-\sigma_A\sqrt{T}$.
Furthermore, the volatility of the
equity is
\begin{equation}
 \sigma_E = \frac{A_t}{E_t} N(d_1) \sigma_A.
\label{sigma_e}%
\end{equation}

The asset value and its volatility, $A_t$ and $\sigma_A$
respectively, cannot be directly observed; however they may be
obtained  solving equations (\ref{merton1}) and (\ref{sigma_e}),
see e.g. \cite{RonnVerma86}. Consequently we can easily  obtain
$d_2$, and
 $PD=Pr[A_T\leq B_T]=N(-d_2)$.

This model has some drawbacks. Its structure implies that  equity
and asset values are non negative in trading markets, whereas
negative asset and equity are possible in accounting, see e.g.
\cite{Peterkort2005}. Furthermore, only part of the total debt is
traded and observable at specific accounting periods. Finally,
Merton's model might underestimate failure probability, see e.g.
\cite{DGFanMa08} and \cite{SuHuang10}. One possible solution to
these issues is proposed in  the following Section.

\subsection{The Default Probability  Model }\label{mod1}

In order to solve the asset observability issue, we
model the firm value via a
 contingent
 claim on the underlying  securities (equity and debt).
We use balance sheet data as a proxy of
 market data, and we apply PCCs to model  the equity dynamic.
For a recent work on  PCCs in contingent claim analysis see \cite{BerCz13}.

 The value of a contingent claim at
 maturity $T$ can be
written in a general form as $G(S_1(T), S_2(T)),$ where $G(\cdot)$
is the pay-off function, and $S_1(T)$ and $S_2(T)$ are the
underlying securities at  maturity $T$. In this framework the
final value of the firm is given by $ \textsf{A}_T =
G(\textsf{E}_T, \textsf{B}_T; T) $
 where
$\textsf{E}_T$ and $\textsf{B}_T$ denote, respectively,   equity
and  debt at maturity $T$. In a similar way we can express the
equity as $ \textsf{E}_T = G_1(\textsf{A}_T, \textsf{B}_T; T) =
(\textsf{A}_T - \textsf{B}_T) $ where $G_1( \cdot)$ is the pay-off
function with density
  $g_1( \cdot)$.
 The equity value at time $t$ is computed as
\begin{equation}\label{equity}
\textsf{E}_t = G_1(\textsf{A}_t, \textsf{B}_t; t) = P(t,T)
\int_{0}^{\infty} \int_{0}^{\infty} G_1(\textsf{A}_T,
\textsf{B}_T; T) g_1(\textsf{A}_T, \textsf{B}_T) d \textsf{A}_T d
\textsf{B}_T,
\end{equation}
where $P(t,T)$ is the risk free discount factor.

The firm value and its return volatility are not directly
observable, hence we use balance sheet data, denoted by $A_T$
(asset) and $B_T$ (liability), as reliable proxy of the market
data, see e.g. \cite{Eberhart05}.
Assets represent what a firm owns, whereas liabilities are debts
 arising from business operations.
% the settlement of
%which is expected to result in an outflow from the firm of
%resources embodying economic benefits.
We decompose $A_T$ and $B_T$ in current ($C_T$) and long term
components ($L_T$) on the basis of the considered time period; that
is $A_T = A_{C_T} + A_{L_T}$ and $B_T = B_{C_T} + B_{L_T}$.
Current assets will be converted into cash within one year,
whereas long term assets will be converted after more than
one year. In a similar way, the firm expects to pay off current liabilities within one year; whereas, the firm expects to settle long term
liabilities after one year.
Comparing current/long term assets with current/long term
liabilities we can obtain a quick gauge of the financial status of
the firm. In fact, standard accounting ratios commonly used  to
investigate the financial strength and efficiency of a firm are
based on these quantities.

 Equation (\ref{equity}) can be rewritten in terms of
balance sheet data as follows

\noindent \begin{eqnarray*}
E_t &=& G_2(A_{C_t}, A_{L_t}, B_{C_t}, B_{L_t}; t) =\\
&=&\! P(t,T) \int_{0}^{\infty}\! \int_{0}^{\infty}\!
\int_{0}^{\infty}\! \int_{0}^{\infty}\! G_2(A_{C_T}, A_{L_T},
B_{C_T}, B_{L_T}; T) \times \\ &\times& g_2(A_{C_T}, A_{L_T},
B_{C_T}, B_{L_T}) dA_{C_T} dA_{L_T} dB_{C_T} dB_{L_T},
\end{eqnarray*}
where $G_2 (\cdot)$ and  $g_2(\cdot)$
are respectively the  pay-off function and its  density for the
decomposed data.

By using Sklar's theorem the 4-dimensional density function \linebreak
$g_2(A_{C_T}, A_{L_T}, B_{C_T}, B_{L_T})$ can be expressed via a copula, and the equity becomes
\begin{eqnarray} E_t &=& P(t,T)
\int_{0}^{\infty} \int_{0}^{\infty} \int_{0}^{\infty}
\int_{0}^{\infty} G_2(A_{C_T}, A_{L_T}, B_{C_T}, B_{L_T}; T) \times  \nonumber\\
 &\times& c(F_{A_C}, F_{A_L}, F_{B_C},
F_{B_L}) f_{A_C} f_{A_L} f_{B_C} f_{B_L} dA_{C_T} dA_{L_T}
dB_{C_T} dB_{L_T},\label{modelEt1}\end{eqnarray}
  where $c(\cdot)$
denotes the $4$-dimensional copula density function, $F_{A_C}$,
$F_{A_L}$, $F_{B_C}$, $F_{B_L}$ are the marginal cumulative
distribution functions, and $f_{A_C}, f_{A_L}, f_{B_C}, f_{B_L}$
are the marginal probability density functions.

In order to improve the flexibility of the model, allowing the
dependence pattern of each pair of variables to be represented by
a different copula, we reexpress the previous equation via
PPCs. The $4$-dimensional copula density function $c(F_{A_C},
F_{A_L}, F_{B_C}, F_{B_L})$ is decomposed in terms of a
sequence of bivariate copulas, not necessary belonging to the same
family of distributions, via a D-vine decomposition.
 The specific decomposition depends
 on the particular data structure under examination; see Section \ref{data_analysis} for the details.

 Simulating from the D-vine decomposition
 we can approximate the equity function in equation (\ref{modelEt1}) via Monte Carlo method as follows
 $$ \tilde{E}_t = P(t,T) \frac{1}{N}
 \sum_{k=1}^N
G_2(\tilde{A}_{C_{ T k}}, \tilde{A}_{L_{T k}}, \tilde{B}_{C_{T
k}}, \tilde{B}_{L_{T k}}; T), $$
 where $N$ is the number of simulations, $\tilde{E}_t,
\tilde{A}_{C_{T k}}, \tilde{A}_{L_{T k}}, \tilde{B}_{C_{T k}}$ and
$\tilde{B}_{L_{T k}}$ are the simulated values of equity, current
and  long term assets and liabilities.  We then estimate  the PD
at time $t$ as $ (PD)_t = Pr(\tilde{E}_t \leq 0)$. More details
about simulating from a D-vine can be found in \cite{AasCzFrBa09}.

\section{Model Estimation}\label{mod_est}

The dynamic of the equity value  in equation   (\ref{modelEt1})
depends on   the parameters of the copula and those of the
marginal distributions. We denote with $\boldsymbol{\theta}$ the
parameter vector of the copula function $c(F_{A_C}, F_{A_L},
F_{B_C}, F_{B_L})$,  and with $\boldsymbol{\delta_m}$ the
parameter vector of the marginal distribution
 $m$, $m \in \{A_C, A_L, B_C, B_L\}$. The vector $\Delta = (\boldsymbol{\delta_{A_C}},
\boldsymbol{\delta_{A_L}}, \boldsymbol{\delta_{B_C}},
\boldsymbol{\delta_{B_L}})$ contains the parameters of  the
marginals, and $\Psi=(\Delta, \boldsymbol{\theta})$ represents the
full set of parameters associated to  (\ref{modelEt1}).

In order to estimate $\Psi$ we follow the {Inference Functions for
Margins} (IFM) procedure proposed by \cite{JoeXu96}.
 The IFM
method estimates the marginal parameters $\Delta$ in a first step,
and then estimates the copula parameters $\boldsymbol{\theta}$, given $\hat{\Delta}_{IFM}$, in a second step.

\subsection{Marginal Parameter Estimation \label{MPE}}

In order to model the marginals we
adopted a parametric approach based on a two-com\-po\-nent Normal
mixture. This approach was motivated by extensive tests and
simulation studies, that are described in Section
\ref{mixture_marg}.

The use of mixture distributions to model multimodal phenomena is
a popular tech\-ni\-que, which has attracted the interest of
several authors in the literature,
see e.g. \cite{McLachPeel00}.
\cite{Peel00} use the ECM
algorithm to fit mixtures of Student's t distributions to data
containing groups of observations with heavy tails or atypical
observations. \cite{KomLes08} propose to model the random effects
of generalised linear mixed models by a mixture of Gaussian
distributions, estimating the parameters in a Bayesian context
using MCMC techniques. \cite{Esco95} use mixture of Dirichlet
processes for density estimation.
For a review of  Bayesian
nonparametric methods for density estimation see e.g.
\cite{Muller2015}.
 \cite{BeChHuYo09} provide a
set of R functions, based on EM algorithms, for analysing a
variety of finite mixture models, such as mixtures of regressions,
multinomial mixtures, nonparametric and semiparametric mixture
models. \cite{SchKau12} represent unknown densities, allowed to
depend on covariates, by a mixture of basis densities, using
penalised splines.

 The current and long term assets
and
 liabilities present  bimodal
 distributions.
 This behaviour can find an explanation in the effect of the managerial actions and decisions performed to
 improve the status of the firm.
 These  actions and decisions directly impact the dynamic of current
and long term assets and
 liabilities, and this can intuitively explain the presence of two separated clusters of data.

Let $F(x_{m_t})$  be the
cumulative distribution function of the marginal $m$ at time $t$.
We estimate each  marginal distribution $F(x_{m_t})$  via a
two-compo\-nent Normal mixture model, assuming different means but
equal variances (location-shift model)
 \begin{equation}
F(x_{m_t}) = \sum_{p=1}^2 \eta_p \Phi(x_{m_t}|\mu_{p},
\sigma^{2}). \label{ls_model}
\end{equation}
In (\ref{ls_model})   $\eta_p$ is the classification probability
for component
 $p$ (with $\eta_p \geq 0$ and $\sum_{p=1}^2 \eta_p = 1$),   and $\Phi(x_{m_t}|\mu_{p},
\sigma^{2})$ is the Normal
 cumulative distribution function with mean $\mu_p$ and variance $\sigma^{2}$.
The likelihood  is given by
$$
L(x_{m}) = \prod_{t=1}^{\mathfrak{n}} \sum_{p=1}^2 \eta_p
\phi(x_{m_t}|\mu_{p}, \sigma^{2}),
$$
\noindent where $\mathfrak{n}$ is the  number of balance sheet
observations, and $\phi$ is the probability density function of
the Normal distribution.

Although based on standard distributions, mixture models pose
highly complex computational challenges. In particular, one major
difficulty is   parameters estimation. The literature about
mixture models offers various solutions both in the classical and
in the Bayesian framework. Considering the classical approach, the
most popular method is the EM algorithm, which is a numerical
optimisation procedure allowing to calculate the maximum
likelihood estimator. However this algorithm may fail to converge
to the
%major
mode of the likelihood, see e.g. \cite{MarMenRob05}. The Bayesian
approach constitutes a more flexible and computationally
convenient solution to the estimation of mixture models, allowing
complex structures to be decomposed into a set of simpler
structures through the use of latent variables. Moreover, the
Bayesian approach permits, via the use of prior distributions, to
incorporate into the model available additional information coming
from different data sources. Furthermore, differently from the
classical approach, the Bayesian one provides reliable parameter
estimates even for sample sizes of limited dimension.

For the previous reasons, we use the  Bayesian approach to model
the dynamic of current and long term asset and liability data. The
posterior distribution of the $m$-th marginal is given by
$$
\pi(\boldsymbol{\delta_m}, \boldsymbol{\eta} |\boldsymbol{x})
\propto \left( \prod_{t=1}^{\mathfrak{n}} \sum_{p=1}^2 \eta_p
\phi(x_{t}|\boldsymbol{\delta_{m}}) \right) \times
\pi(\boldsymbol{\delta_m}, \boldsymbol{\eta}),
$$
where $\boldsymbol{x}$ is the
balance sheet data vector, $\pi(\boldsymbol{\delta_m},
\boldsymbol{\eta})$ is the joint prior distribution of
$\boldsymbol{\delta_m}$ and the vector of classification
probabilities $\boldsymbol{\eta}$. The posterior
$\pi(\boldsymbol{\delta_m}, \boldsymbol{\eta} |\boldsymbol{x})$ is
computationally intractable
 to work with; hence, the data augmentation MCMC algorithm is used to estimate the parameters of the mixture
 distributions, see \cite{TanWong87}.
 The data
augmentation algorithm introduces a vector of latent variables
$\boldsymbol{z}=(z_1, \ldots, z_ \mathfrak{n})$, that represents
the allocations associated to each observation $x_t$. Hence, the
posterior density can be expressed as
$$
\pi(\boldsymbol{\delta_m}, \boldsymbol{\eta} |\boldsymbol{x}) =
\int_Z \pi(\boldsymbol{\delta_m}, \boldsymbol{\eta}
|\boldsymbol{z}, \boldsymbol{x}) \pi(\boldsymbol{z}|
\boldsymbol{x}) d\boldsymbol{z},
$$
where $\pi(\boldsymbol{z}| \boldsymbol{x})$ denotes the predictive
density of the latent data $\boldsymbol{z}$ given
$\boldsymbol{x}$, with $\boldsymbol{z}=(z_1, \ldots,
z_{\mathfrak{n}})$, and $\pi(\boldsymbol{\delta_m},
\boldsymbol{\eta} |\boldsymbol{z}, \boldsymbol{x})$ is the
conditional density of the parameters given the augmented data.
Moreover, $\pi(\boldsymbol{\delta_m}, \boldsymbol{\eta}
|\boldsymbol{z}, \boldsymbol{x}) = \pi(\boldsymbol{\delta_m}|
\boldsymbol{\eta}, \boldsymbol{z},
\boldsymbol{x})\pi(\boldsymbol{\eta} |\boldsymbol{z},
\boldsymbol{x})$, and $\pi (\boldsymbol{\eta} |\boldsymbol{z},
\boldsymbol{x}) = \pi (\boldsymbol{\eta} |\boldsymbol{z})$, since
the distribution is independent of $\boldsymbol{x}$. The data
augmentation algorithm uses an iterative procedure simulating
$\boldsymbol{z}$ first, then generating $\boldsymbol{\eta}$ from
$\pi (\boldsymbol{\eta} |\boldsymbol{z})$ and finally generating
$\boldsymbol{\delta_m}$ from $\pi (\boldsymbol{\delta_m}
|\boldsymbol{\eta}, \boldsymbol{z}, \boldsymbol{x})$. %This
%approach is motivated by the fact that
The densities $\pi
(\boldsymbol{\eta} |\boldsymbol{z})$ and $\pi
(\boldsymbol{\delta_m} |\boldsymbol{\eta}, \boldsymbol{z},
\boldsymbol{x})$, are easier to sample than the original
posterior.

Assuming independency between parameters {a priori}, we specify
the following prior distributions
\begin{eqnarray*}
z_t &\sim& {Bernoulli} ( \eta_1)\\
(\eta_1, \eta_2) &\sim& Dirichlet(\alpha_1, \alpha_2)\\
\mu_p &\sim& Normal(b_p, B_p)\\
\sigma^{2} &\sim& \Gamma^{-1} \left( \nu/2, \nu S /2  \right),
\end{eqnarray*}
where a convenient choice of hyperparameters $\alpha_1, \alpha_2,
b_p, B_p, \nu, S$ leads us to vague prior distributions. A
sensitivity analysis was carried out proposing different
hyperparameter values; however, the high similarity of all results
suggested that the model is insensitive to prior parameter choice.

We need to point out that the simulations were implemented using the software JAGS (Just Another Gibbs Sampler; \cite{Plummer03}),
where the risk of unidentifiability of the model due to label switching was avoided specifying the
constraint of unique ordering of the segments, with ascending
means of the segment distributions.

\subsection{Copula Parameter Estimation \label{CPE}}

 To estimate the copula parameters
$\boldsymbol{\theta}$ we apply the following  five phases
procedure.

In the first phase a suitable D-vine decomposition is selected to
model the copula  $c(F_{A_C}, F_{A_L}, F_{B_C}, F_{B_L};
\boldsymbol{\theta})$.  We select as a first tree the one
maximizing the pairwise dependencies between the considered
variables. As a measure of pairwise dependence we use the
Kendall's $\tau$, calculated for each edge connecting two nodes.
The problem of finding the maximum weighted sequence of the
variables can be transformed into a travelling salesman problem
instance and solved accordingly, see \cite{Brechmann}. The
structure of remaining trees is completely determined by the
structure of the first one.
 Therefore, the
strongest dependencies are captured in the first tree, allowing to
obtain a more parsimonious model, with more stable parameter
estimates.

In the second phase suitable pair copulas are chosen. For each
pair of variables we select the best fitting pair copula using the
Akaike Information Criterion (AIC), which is chosen among other
criteria (i.e. the Vuong and Clarke goodness-of-fit test developed
by \cite{Vuong89} and \cite{Clarke07}, and Bayesian Information
Criterion (BIC)) for its good performance in simulation studies.
However, before calculating the AIC, the Genest and Favre
 bivariate asymptotic independence test (\cite{GenFav07}) is performed to check for independence on each pair
  of variables of the D-vine. If conditional independence between variables is observed,
the number of levels of the pair copula decomposition is reduced, and hence
the construction is simplified, as discussed in Section \ref{Dvine}.

In the third phase, the parameters of the copulas in the first
tree are estimated. For each copula there is at least one
parameter to be determined. The number of parameters depends on
which copula type is selected in the previous phase. To estimate
the copula parameters we employ the maximum likelihood estimation
method, using the sequential updating parameter estimates as
starting values, see \cite{AasCzFrBa09} for more details.

In the fourth phase, given the results of the first tree, we compute pseudo-observations via the
conditional distributions $F(x|\textbf{v})$. These values are then
used as input for the next trees of the D-vine.

In the fifth phase, the procedure illustrated from phase $2$ to
phase $4$ is repeated for all trees of the D-vine.

\section{Empirical Analysis}\label{data_analysis}

%For comparative purpose, to test the behaviour of our methodology,
%we also examined a non defaulted firm with a strong financial
%reputation, and  presenting some characteristics in common with
% the two of defaulted examined firms (Cirio and Parmalat)
%
%We apply our methodology to the analysis of both defaulted and
%operative firms.

We consider four fraudulent bankruptcy cases, related to well
known   financial scandals: Cirio (1993-2002), Enron (1997-2000),
Parmalat (1990-2003), Swissair (1988-2000). To test the behaviour
of our methodology, we also examine the Sysco company, a firm
operating in the same period of time of the previous ones, with a
strong financial reputation, and presenting some characteristics
in common with some of the examined defaulted firms. For
comparative purposes, for this last firm we consider balance sheet
data of the  years 1990-2003. With the exception of Enron, the
other defaulted firms are now operating under the direction of a
different leadership group.

 We use semestral balance
sheets data downloaded by the ``Thomson Reuters'' and the
``Bloomberg'' databases. The data have been converted into monthly
observations assuming uniform distribution in the semesters. For
Swissair and Enron the complete balance sheets for the year of
failure are not available.

We now briefly describe the profile of each examined  firm,
outlying the events that lead to the
 bankruptcy of the defaulted firms.

 Cirio is an Italian food company founded in 1856. Its bankruptcy in 2002 was the consequence of the
fraudulent financial policy of its managerial group.

Enron was an American energy, commodities, and services company
created in
 1985 through the merger of two natural gas companies.
Before its collapse in 2001 it was one of America's leading
companies with a solid reputation, and it was one of the highest-rated
companies of Wall Street.
 At the end of 2001 it was made
 public that its apparently solid financial conditions were substantially sustained by an
institutionalised, systematic, accounting fraud. The company
declared bankruptcy in December 2001.

Parmalat was created in  1961 as a small pasteurisation plant in
Parma (Italy). It subsequently became a multinational corporation
in the 80's with different food product lines, and expanded
further in the 90s. It was listed for the first time in the Milan
stock exchange in 1990.
 Parmalat collapse in 2003 was the biggest
case of financial fraud and money laundering perpetrated by a
private company in Europe.
 It was the first Italian corporate
crash with  international implications.

Swissair presents a different story from the previous defaulted
firms. It was formed in 1931 from the merging between Balair and
Ad Astra Aero and it was one of the major international
airlines with a strong financial stability. It rapidly declined
from one of the major international airlines with the strongest
balance  into bankruptcy in 2001. This rapid decline was the
consequence of inefficient alliance policies, management inability
and economic turndown following the terroristic attacks of
``September 11''.

Sysco is an American marketer and distributor of foodservice
products. It was founded in 1969 and became public in 1970.
Nowadays, it is  a solid company with a very good reputation.

In the following Section we report a detailed analysis of the four
defaulted companies, and  we present the main important results of
the Sysco company.

 \subsection{Mixture Models for Asset and Liability Data}\label{mixture_marg}

We modeled the current and long
term assets and liabilities of the considered companies employing
the two-component Normal mixture model described in Section
\ref{MPE}. The choice of this model was determined by extensive
tests and simulation studies.

First, in order to identify the best model for the marginals,
we assessed the fit to the data of classical parametric models, such as the Normal, the left-truncated
 Normal in zero, the log-Normal, the Gamma, the Exponential and the Weibull distribution.
 We implemented a bootstrap version of the univariate Kolmogorov-Smirnov test, with 1,000 Monte Carlo simulations.
  The bootstrap Kolmogorov-Smirnov tests for the hypothesis that the actual data were generated by the corresponding theoretical distribution.
Table \ref{KStest} shows the results of the bootstrap
Kolmogorov-Smirnov tests for the marginals of the Enron dataset.
We obtained very similar outputs for the other datasets considered
in this paper. For each theoretical distribution being tested the
average p-value over the 1,000 simulations and the percentage of
times the null hypothesis is not rejected are displayed.
 Since the null hypothesis was always rejected at the 0.05 level,
 we concluded that none of the classical parametric model tested was suitable for our data.

\begin{table}[htbp]
\caption{Bootstrap Kolmogorov-Smirnov tests for the marginals of the Enron dataset. The columns list the four marginals and the rows display the average p-values and the percentages of times when the null hypothesis is not rejected, for the six classical parametric models considered.}\label{KStest}%
 \vspace{0.3cm}
  \centering
  \begin{footnotesize}
    \begin{tabular}{cccccc}
    \hline
    \textbf{Distributions} & \textbf{Marginals} & \textbf{$A_{C_{T}}$}   & \textbf{$A_{L_{T}}$}   & \textbf{$B_{C_{T}}$}   & \textbf{$B_{L_{T}}$} \\
    \hline
    \textbf{Normal} & \textbf{Average p-value} & 0.00000 & 0.00032 & 0.00000 & 0.00002 \\
          & \textbf{\% non-rejected $H_0$} & 0.00  & 0.00  & 0.00  & 0.00 \\
    \textbf{Left-Truncated Normal in 0} & \textbf{Average p-value} & 0.00000 & 0.00055 & 0.00000 & 0.00000 \\
          & \textbf{\% non-rejected $H_0$} & 0.00  & 0.00  & 0.00  & 0.00 \\
    \textbf{Log-Normal} & \textbf{Average p-value} & 0.00425 & 0.00173 & 0.00184 & 0.00021 \\
          & \textbf{\% non-rejected $H_0$} & 0.01  & 0.00  & 0.00  & 0.00 \\
    \textbf{Gamma} & \textbf{Average p-value} & 0.00000 & 0.00157 & 0.00000 & 0.00004 \\
          & \textbf{\% non-rejected $H_0$} & 0.00  & 0.00  & 0.00  & 0.00 \\
    \textbf{Exponential} & \textbf{Average p-value} & 0.00000 & 0.00000 & 0.00000 & 0.00000 \\
          & \textbf{\% non-rejected $H_0$} & 0.00  & 0.00  & 0.00  & 0.00 \\
    \textbf{Weibull} & \textbf{Average p-value} & 0.00000 & 0.00023 & 0.00000 & 0.00000 \\
          & \textbf{\% non-rejected $H_0$} & 0.00  & 0.00  & 0.00  & 0.00 \\
    \hline
    \end{tabular}%
    \end{footnotesize}
\end{table}%

\begin{figure}[htbp]
\begin{center}
 \includegraphics[width=12cm]{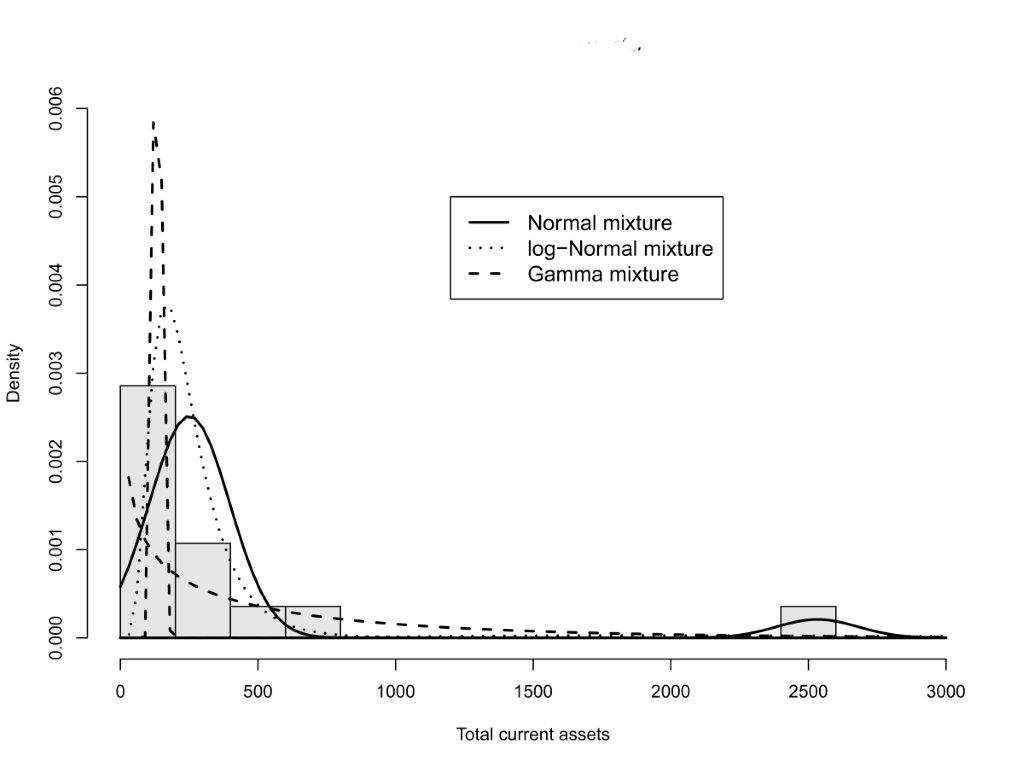}
 \caption{Histogram of Enron current assets (in EUR) fitted with Normal (solid line), log-Normal (dotted line), and Gamma (dashed line) mixture models.}\label{other_mixtures}
\end{center}
\end{figure}

Due to the poor fit of classical models to our marginals,
and since the current and long term assets and liabilities present bimodal distributions,
we opted for two-component parametric mixture models. We tested several families of parametric
 mixture distributions for the marginals of all the considered datasets, and the Normal mixture
  always outperformed other models. In particular, we selected the Normal, the log-Normal and the
  Gamma mixtures, since many other models are related to them: the truncated Normal is related to
  the Normal and the Exponential and Weibull are related to the Gamma distribution.
  Figure \ref{other_mixtures} depicts the histogram of Enron current assets (in EUR) fitted with
   three different mixture models and the Normal mixture (solid line) clearly shows the best fit.
    Similar results were obtained for the remaining marginals and datasets.

Before choosing to model the marginals with a two-component Normal
mixture model, we estimated  the number of components $p$ using
Bayes factors, as suggested by \cite{KassRaf95},
\cite{RichGreen97},  and \cite{MarMenRob05}.
%Bayes factor is a model selection method based on the ratio
%between the probabilities that the data are produced by two
%competing models.
 We calculated Bayes factors for all the
marginals of the considered companies, comparing the model with
two components with all models with a number of components $p=1,
3, 4, \ldots, 10$. Placing the model with two components in the
numerator of the Bayes factors, the results we obtained were
greater than one for all the marginals, showing that the
two-component Normal mixture is the most strongly supported model
by the data. For illustration, Table \ref{Bayesfactor} lists the
Bayes factor results for the Enron current assets data. $BF_{r,s}$
denotes the Bayes factor of model $r$ against model $s$. The
results are not surprising, since inspection of the histograms of
the marginals clearly reveals bi-modal distributions.

\begin{table}[htbp]
\caption{Bayes factors for Enron current assets data. Each Bayes
factor compares the model with two components against models with
a
 number of components $p=1, 3, 4, \ldots, 10$.}\label{Bayesfactor}%
 \vspace{0.3cm}
  \centering
    \begin{tabular}{cc}
    \hline
    \textbf{Competing models} & \textbf{Bayes factor} \\
    \hline
    $BF_{2,1}$ & 2.7979 \\
    $BF_{2,3}$ & 2.8601 \\
    $BF_{2,4}$ & 3.0305 \\
    $BF_{2,5}$ & 3.1008 \\
    $BF_{2,6}$ & 3.1558 \\
    $BF_{2,7}$ & 3.1926 \\
    $BF_{2,8}$ & 3.4840 \\
    $BF_{2,9}$ & 2.9698 \\
    $BF_{2,10}$ & 2.9564 \\
    \hline
    \end{tabular}%
  \end{table}

\begin{figure}[htbp]
\begin{center}
 \includegraphics[width=11cm]{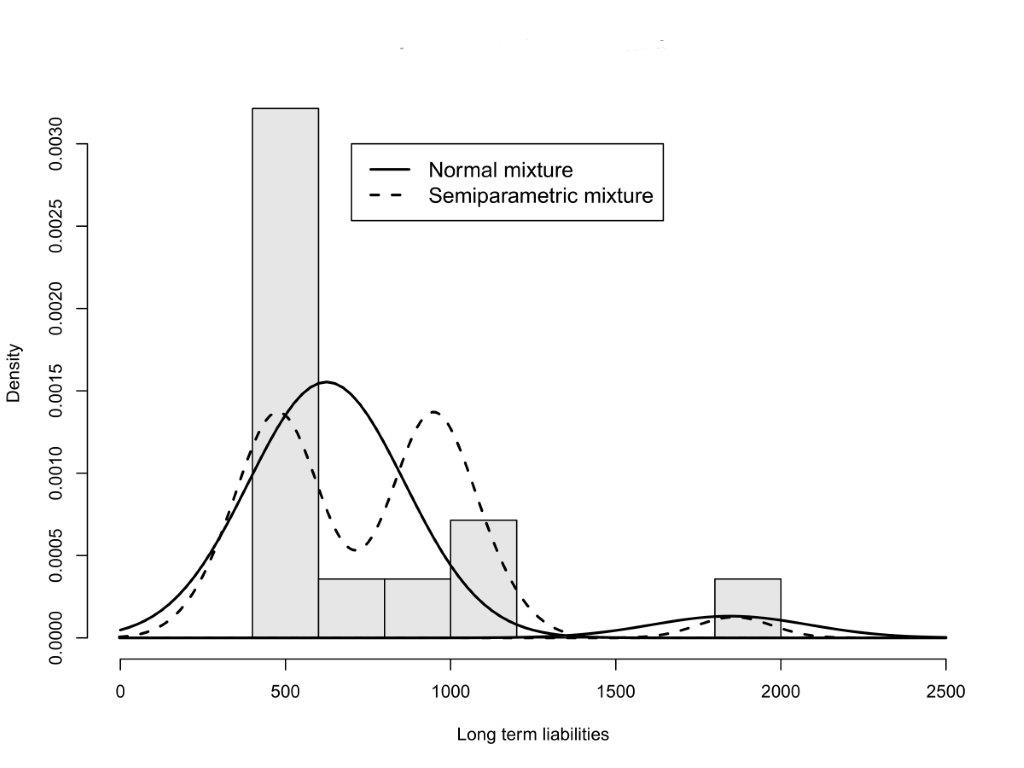}
   \includegraphics[width=11cm]{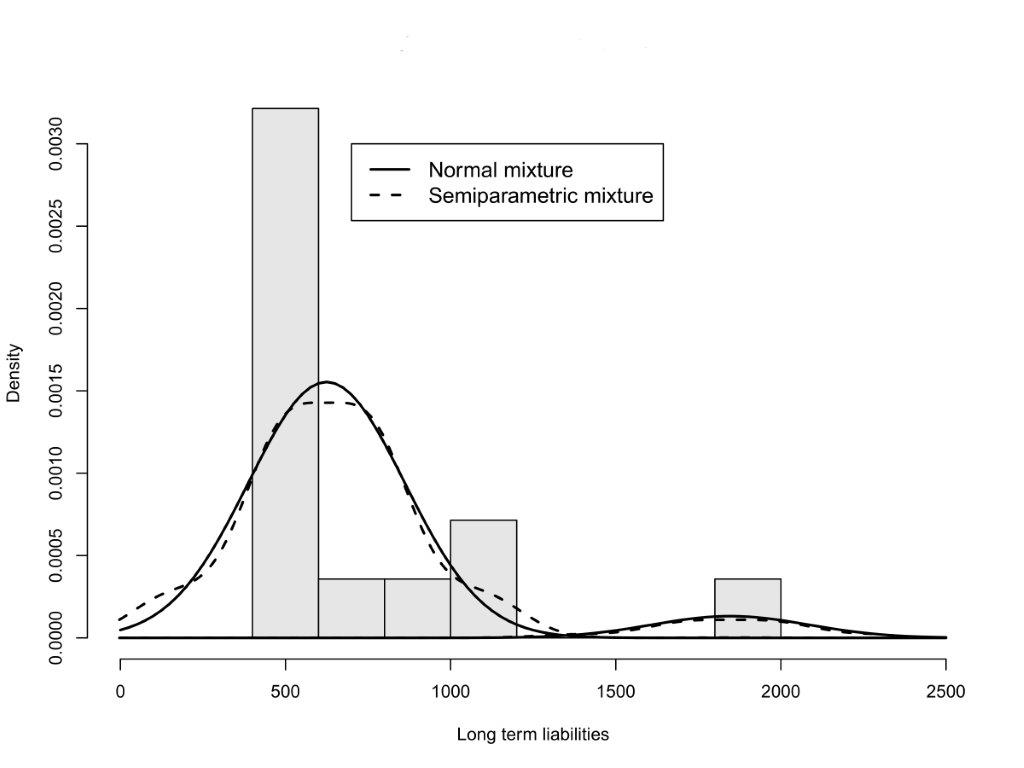}
  \caption{Histogram of Enron long term liabilities (in EUR) fitted
 with the 2-components Normal mixture (solid line) and the semiparametric model (dashed line). The semiparametric model was estimated with
 (second row) and without (first row) specifying the initial mean values of the mixture components.}\label{semiparametric}
\end{center}
\end{figure}

In addition, we tested the fit
of the two-component Normal mixture with the symmetric
location-shifted semiparametric model of \cite{BoMoVa06} and
 \cite{HuWaHe07}, which is based on a mixture of unspecified
densities, assumed symmetric about zero, see \cite{BeChHuYo09}.
Figure \ref{semiparametric} shows the histogram of Enron long term
liabilities (in EUR) fitted with the two-components Normal mixture
(solid line) and the semiparametric model (dashed line). In the
first row plot the semiparametric model was obtained with no
specification of the initial mean values of the mixture
components, while in the second row plot the semiparametric model
was obtained specifying the initial mean values of the mixture
components. In the former case, the semiparametric model shows a
worse performance than the Normal mixture, adding a new
unnecessary component to the mixture. In the latter case the
performance of the semiparametric model is very similar to the
Normal mixture and does not improve the fit to the original data.
The application of the semiparametric model to the remaining
marginals of the other considered datasets yields very similar
results.

Therefore, we modelled the current
and long term assets and liabilities using a two-component Normal
mixture, since this was the best model to fit the marginals. For
each single firm we report the estimates of the parameters (posterior means) of the
corresponding mixture models in Table \ref{mixture_pars}, together with the 95\% credible intervals (in brackets).

\begin{sidewaystable}
%\begin{table}[htbp]
\caption{Parameter estimates (posterior means) of the marginal distributions. The 95\% credible intervals are in brackets.}\label{mixture_pars}%
 \vspace{0.3cm}
  \centering
  \begin{footnotesize}
 % \begin{scriptsize}
    \begin{tabular}{cccccc}
    \hline
    \textbf{Cirio} & \textbf{$\eta_1$} & \textbf{$\eta_2$} & \textbf{$\mu_1$} & \textbf{$\mu_2$} & \textbf{$\sigma^2$} \\
    \hline
    $F_{A_C}$   & 0.635 (0.5386; 0.7241)  & 0.365 (0.2759; 0.4614) & 40.28 (35.78;  44.56) & 111.68 (105.93; 117.92) & 272.94 (207.36; 360.09) \\
    $F_{A_L}$   & 0.5096 (0.4156; 0.6009) & 0.4904 (0.3991; 0.5844) & 18.95 (17.51; 20.38) & 37.84 (36.40; 39.24) & 26.261 (19.19; 36.50) \\
    $F_{B_C}$   & 0.5012 (0.4091; 0.5923) & 0.4988 (0.4077; 0.5909) & 39.54 (36.42;  42.68) & 106.43 (103.16; 109.57) & 163.79 (126.2; 212.4) \\
    $F_{B_L}$   & 0.6954 (0.6101; 0.7757) & 0.3046 (0.2243; 0.3899) & 16.7 (14.52; 18.89) & 67.24 (63.71; 70.56) & 100.82 (77.44; 130.77) \\
    \hline
          &       &       &       &       &  \\
    \hline
    \textbf{Enron} & \textbf{$\eta_1$} & \textbf{$\eta_2$} & \textbf{$\mu_1$} & \textbf{$\mu_2$} & \textbf{$\sigma^2$} \\
    \hline
    $F_{A_C}$   & 0.92336 (0.88023; 0.9598)  & 0.07664 (0.04022; 0.1198) & 251.3 (227.7;  274.5)  & 2531.6 (2454.8; 2612.7) & 21412 (17207; 26786) \\
    $F_{A_L}$   & 0.7133 (0.6439; 0.7791) & 0.2867 (0.2209; 0.3561) & 556.9 (547.6; 566.3) & 880.7 (865.5; 896.4) & 2834.1 (2267.8; 3532.1) \\
    $F_{B_C}$   & 0.92334 (0.87994; 0.9586) & 0.07666 (0.04143; 0.1201) & 260.7 (239.9; 280.7)  & 2367.4 (2296.5; 2438.5) & 17062 (13768; 21157) \\
    $F_{B_L}$   & 0.92231 (0.87554; 0.9576)  & 0.07769 (0.04236; 0.1245) & 623.7 (588.7; 660.5) & 1843.9 (1693.9; 1990.2)  & 56107 (44994; 69076) \\
    \hline
          &       &       &       &       &  \\
     \hline
    \textbf{Parmalat} & \textbf{$\eta_1$} & \textbf{$\eta_2$} & \textbf{$\mu_1$} & \textbf{$\mu_2$} & \textbf{$\sigma^2$} \\
    \hline
    $F_{A_C}$   & 0.615 (0.5396; 0.6893) & 0.385 (0.3107; 0.4604) & 121.9 (109.2; 134.5) & 386.3 (369.7; 403.4) & 3852.4 (3044.4; 4815.9) \\
    $F_{A_L}$   & 0.5375 (0.4658; 0.6084) & 0.4625 (0.3916; 0.5342) & 63.7 (58.85; 68.43) & 175.2 (170.14; 180.50) & 535.47 (425.71; 666.26) \\
    $F_{B_C}$   & 0.92823 (0.88568; 0.9618) & 0.07177 (0.03821; 0.1143) & 194.3 (170.9; 217.2) & 1587.8 (1500.4; 1674.0) & 23362 (18985; 28988) \\
    $F_{B_L}$   & 0.9278 (0.8882; 0.9616) & 0.0722 (0.0384; 0.1118) & 194 (170.3; 216.3) & 1588.8 (1500.4; 1675.5) & 23228 (19097; 28716) \\
    \hline
          &       &       &       &       &  \\
    \hline
    \textbf{Swissair} & \textbf{$\eta_1$} & \textbf{$\eta_2$} & \textbf{$\mu_1$} & \textbf{$\mu_2$} & \textbf{$\sigma^2$} \\
    \hline
    $F_{A_C}$   & 0.5442 (0.4612; 0.6216) & 0.4558 (0.3784; 0.5388) & 162.5 (154.5; 171.8) & 297.1 (288.6; 306.2) & 1214.3 (923.6; 1609.7) \\
    $F_{A_L}$   & 0.2339 (0.1696; 0.3034) & 0.7661 (0.6966; 0.8304) & 90.83 (79.09; 103.1) & 349.66 (343.10; 356.3) & 1373.1 (1104.6; 1731.0) \\
    $F_{B_C}$   & 0.91266 (0.86004; 0.9524) & 0.08734 (0.04762; 0.1400) & 176.4 (169.4; 183.6) & 350.5 (319.9; 378.6) & 1901.9 (1502.5; 2405.3) \\
    $F_{B_L}$   & 0.5402 (0.4606; 0.6183) & 0.4598 (0.3817; 0.5394) & 163.4 (148.8; 177.8) & 439.5 (424.7; 455.0) & 4259 (3345; 5412) \\
    \hline
       &       &       &       &       &  \\
    \hline
    \textbf{Sysco} & \textbf{$\eta_1$} & \textbf{$\eta_2$} & \textbf{$\mu_1$} & \textbf{$\mu_2$} & \textbf{$\sigma^2$} \\
    \hline
    $F_{A_C}$   & 0.562 (0.4974; 0.6262) & 0.438 (0.3738; 0.5026) & 347.9 (323.5; 372.7) & 811.2 (782.5; 839.5) & 16877 (13847; 20727) \\
    $F_{A_L}$   & 0.5649 (0.5061; 0.6248) & 0.4351 (0.3752; 0.4939) & 272.3 (251.9; 291.7) & 830.9 (807.2; 854.0) & 14281 (11983; 17031) \\
    $F_{B_C}$   & 0.5005 (0.4401; 0.5637) & 0.4995 (0.4363; 0.5599) & 184.3 (171.8; 197.0) & 518.0 (504.7; 531.2) & 4982.4 (4177.6; 5903.8) \\
    $F_{B_L}$   & 0.6356 (0.5760; 0.6945) & 0.3644 (0.3055; 0.4240) & 187.9  (174.2; 201.3) & 546.3 (526.0; 565.4) & 6875.2  (5765.7; 8190.0) \\
    \hline
    \end{tabular}%
 %   \end{scriptsize}
\end{footnotesize}
%  \end{table}%
\end{sidewaystable}

The classification probabilities $\eta_p$ are quite close to $0.5$
for Cirio data, for the asset marginals of Parmalat data,   for
the current assets and long term liabilities of Swissair data, and
for Sysco data, denoting a balanced number of observations in the
two mixture components. On the contrary, Enron data, the liability
marginals of Parmalat data, long term assets and current
liabilities of Swissair data show  very different classification
probabilities $\eta_1$ and $\eta_2$. This means that different
proportions of observations are allocated to the  components of
the mixture and that one of the two components captures the
greatest number of data. The location parameters of the two Normal
components of the mixture $\mu_p$ are well separated, especially
for Enron, Parmalat and Sysco, denoting that the mixture model is
able to express the mean difference between the two components.
The dispersion parameter $\sigma^2$ is particularly high for
Enron, Parmalat and Sysco,
 while it is lower for Cirio and Swissair.

Enron and Parmalat have the most
 unbalanced mixture components, especially with reference to the liability marginal data.
 The data of these two companies are
characterised by very different values of classification
probabilities $\eta_p$, very different  means $\mu_p$, and very
high normal variance values $\sigma^2$. The resemblance of the
structure of assets and liabilities in Enron and Parmalat may be
explained by the similar behavior of these two companies during
the years before their default. Parmalat indeed has been referred
to as the ``Europe's Enron''.

We now present  a  graphical analysis of the results of Enron
company. We have performed a similar analysis for the remaining
four companies, but we do not report it here for lack of space.

\begin{figure}
\begin{minipage}{0.5\linewidth}
 \includegraphics[width=9cm]{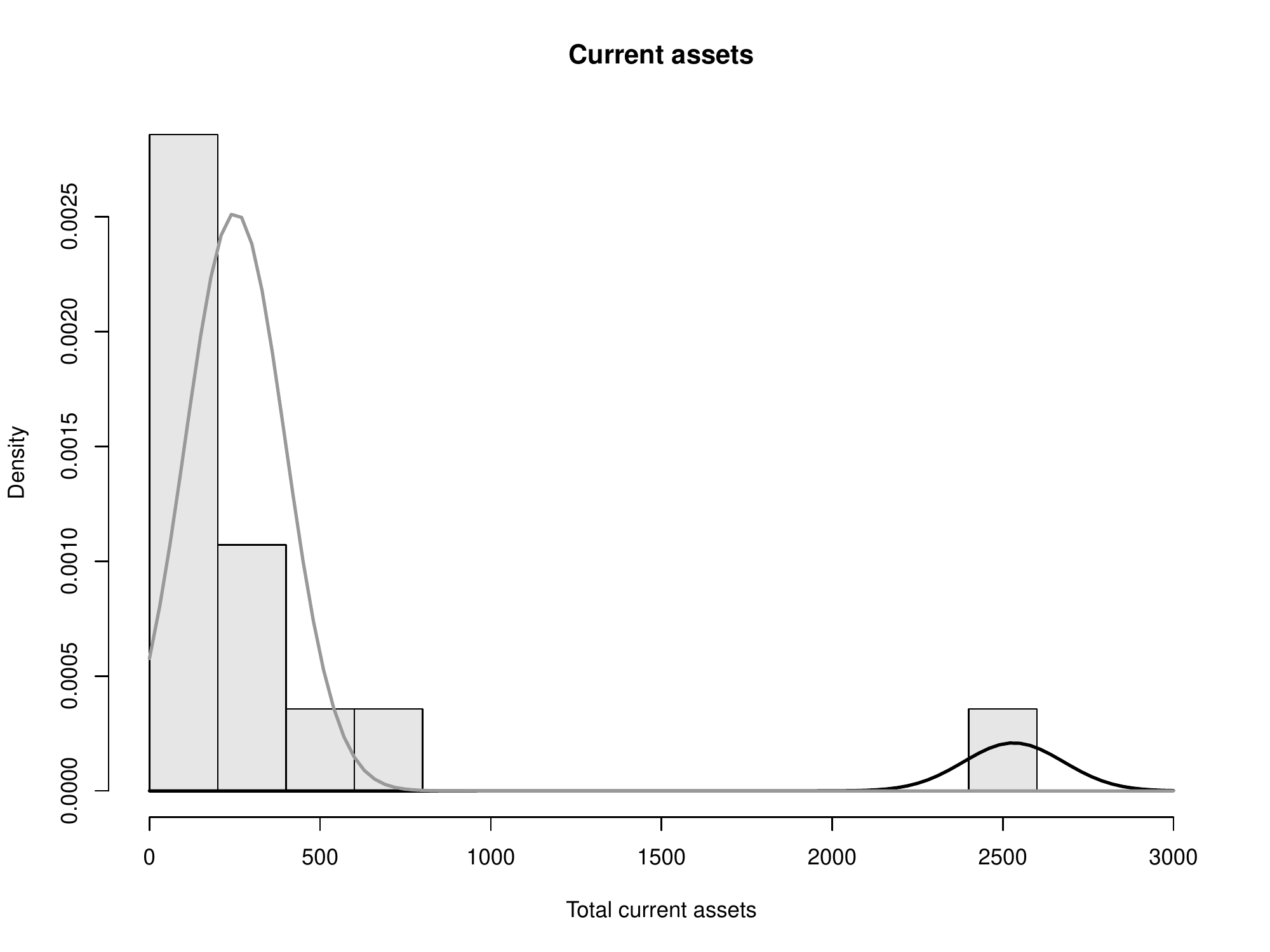}
 \end{minipage}
\begin{minipage}{0.5\linewidth}
\includegraphics[width=9cm]{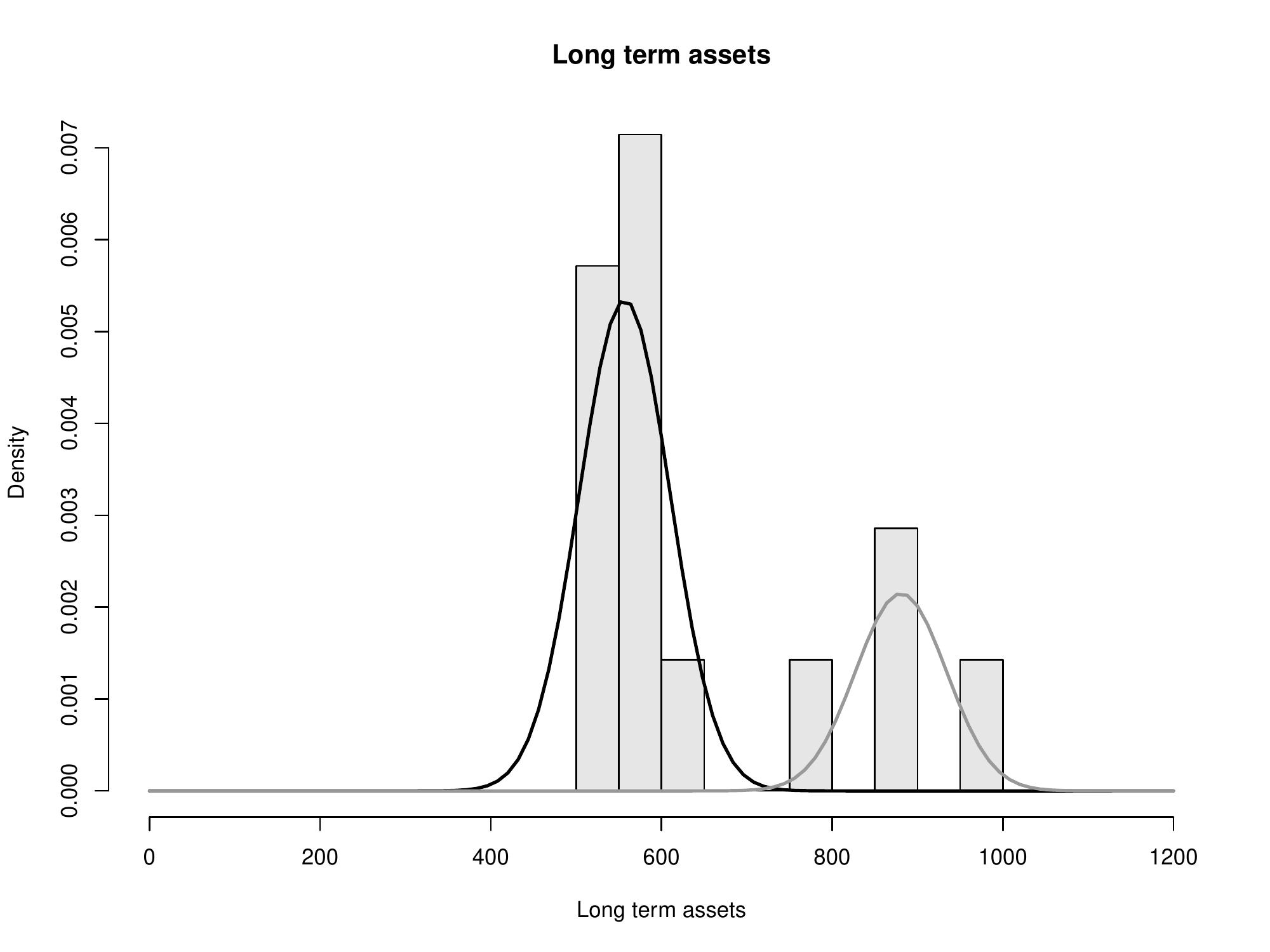}
\end{minipage}\\
\begin{minipage}{0.5\linewidth}
 \includegraphics[width=9cm]{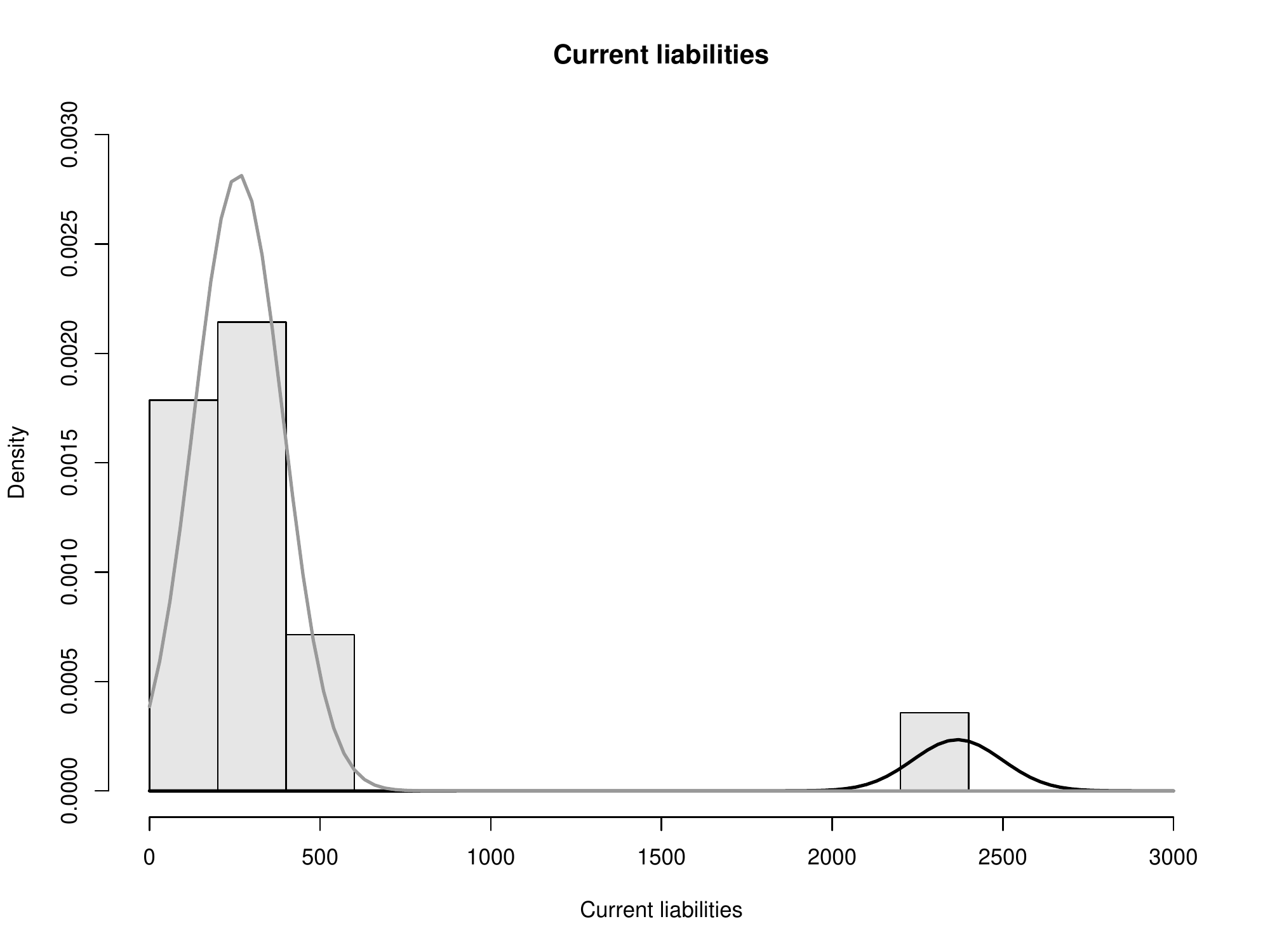}
\end{minipage}
\begin{minipage}{0.5\linewidth}
  \includegraphics[width=9cm]{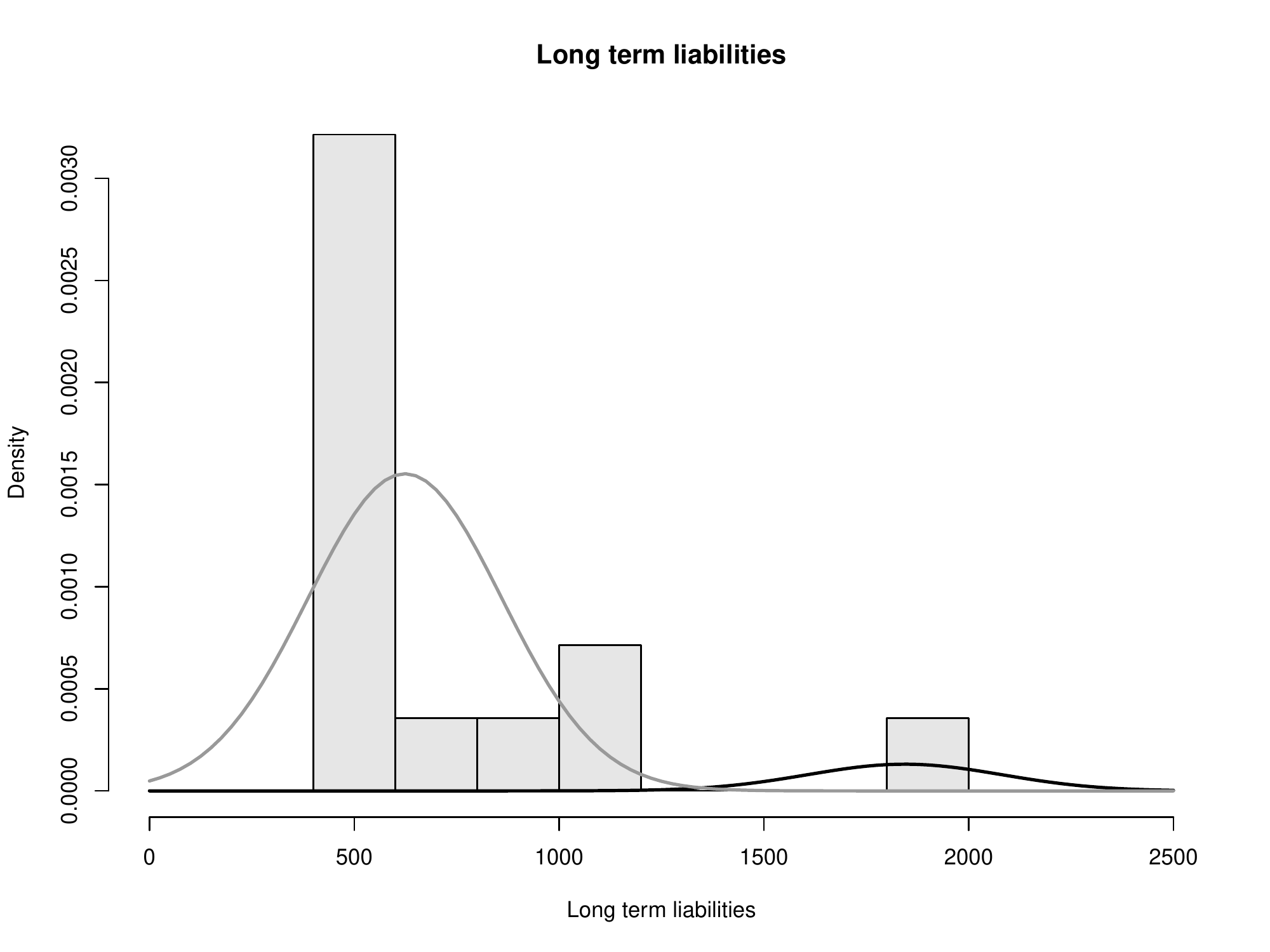}
  \end{minipage}
\caption{Enron data, measured in EUR, fitted with a mixture of two Normal
components: $F_{A_C}$ (top left), $F_{A_L}$ (top right), $F_{B_C}$
(bottom left) and $F_{B_L}$ (bottom right).}\label{enron_hist}
\end{figure}

Figure \ref{enron_hist} shows the histograms of each marginal measured in EUR
(grey bars), fitted with the location-shift model of two Normal
components (black  and grey lines) described in Section \ref{MPE}.
$F_{A_C}$ is displayed in the top left
panel, $F_{A_L}$ in the top right, $F_{B_C}$ in the bottom left
and $F_{B_L}$ in the bottom right. Let us consider the picture
related to the  current assets marginal of Enron data (top left
panel of Figure \ref{enron_hist}). The histogram shows a highly
bimodal distribution which justifies the use of a finite mixture
model. Similar comments arise from the analysis of the fitted
histograms of the remaining marginals.

\begin{figure}
 \includegraphics[width=22cm]{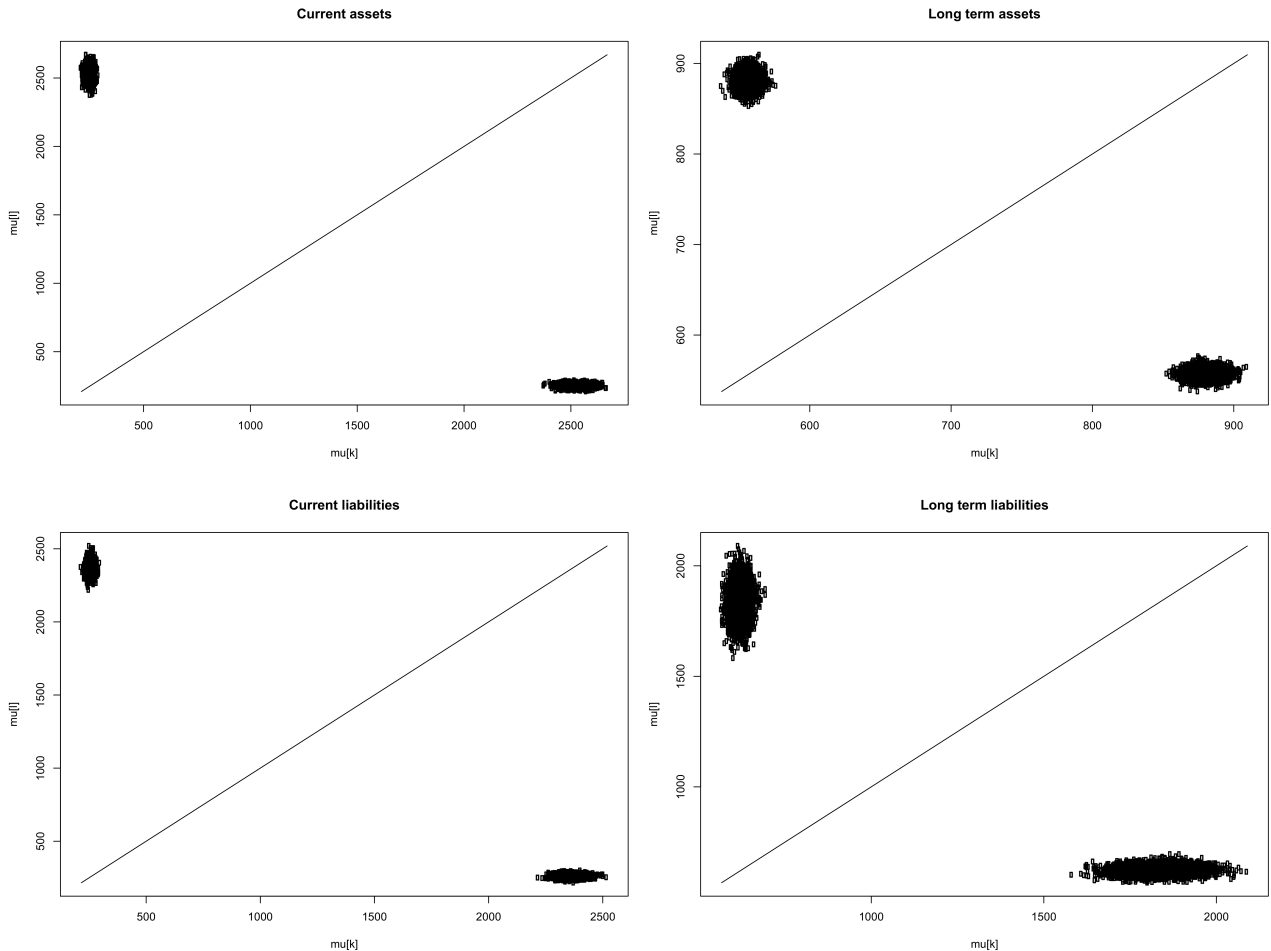}
\vspace{-14.5cm} \caption{Enron data: $\mu_1$ versus $\mu_2$.
$F_{A_C}$ is in the top left, $F_{A_L}$ is in the top right,
$F_{B_C}$ is in the bottom left and $F_{B_L}$ is displayed in the
bottom right panel.}\label{enron_mu1vsmu2}
\end{figure}

Figure \ref{enron_mu1vsmu2} shows the sampled values of the
$\mu_1$ parameter on the horizontal axis and of the $\mu_2$
parameter on the vertical axis.
$F_{A_C}$ is displayed in top left panel, $F_{A_L}$ in the top
right, $F_{B_C}$ in the bottom left and $F_{B_L}$ in the bottom
right. It is interesting to note that our data are not affected by
label switching, since the segments are rather well separated for
$\mu$, as there are no points on the diagonal on the $\mu_1$
versus $\mu_2$ plots.

\begin{figure}[htbp]
\begin{center}
  \includegraphics[width=24cm]{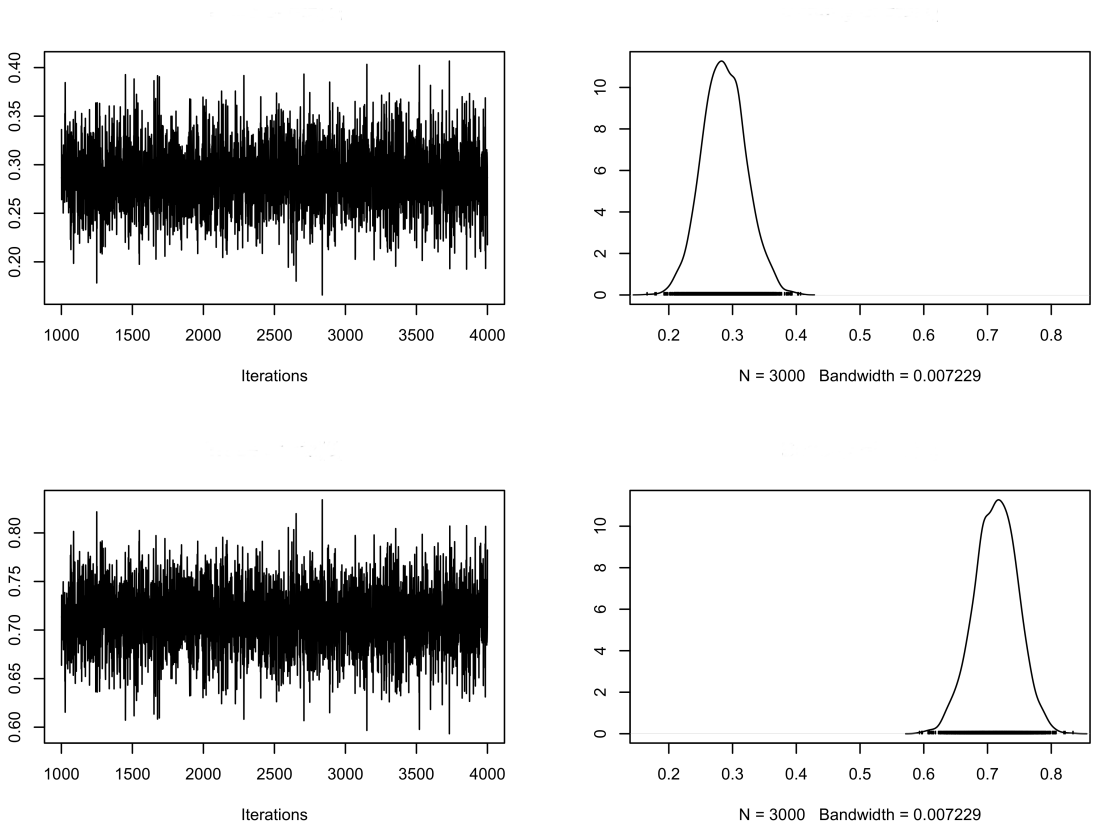}
 \vspace{-20cm}\caption{Enron long term assets data: MCMC traces (left panels) and posterior densities (right panels) for $\eta$; with $\eta_1$ on the first row and
 $\eta_2$ on the second row.}\label{enron_fa_pars_1}
\end{center}
\end{figure}

\begin{figure}[htbp]
\begin{center}
  \includegraphics[width=22.5cm]{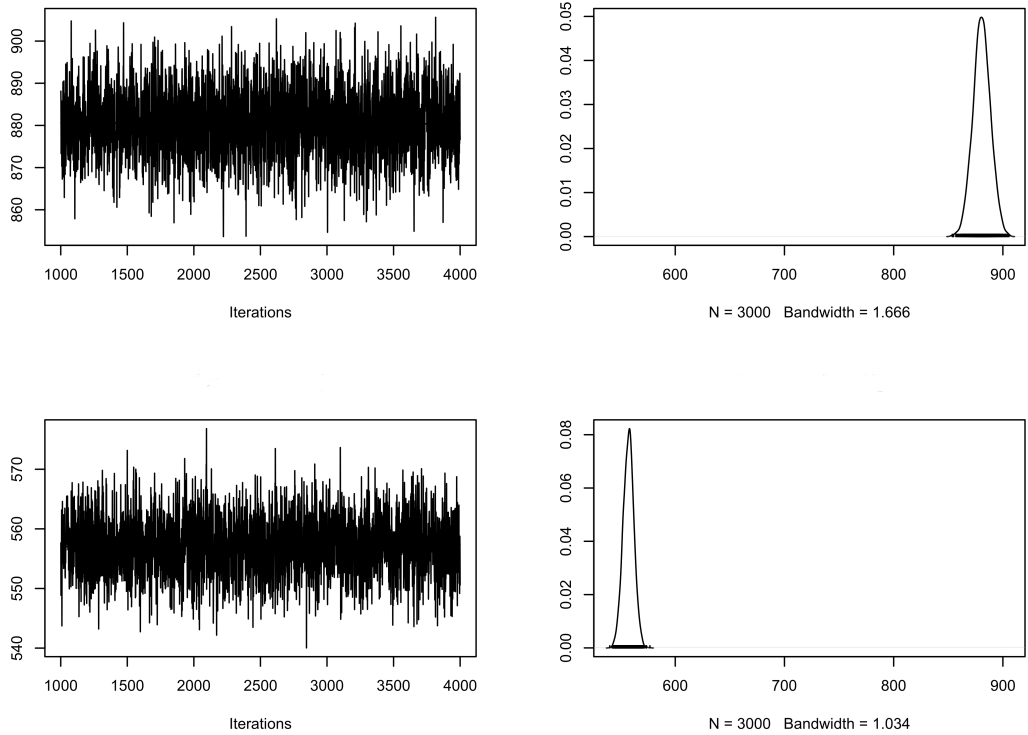}\\
  \vspace{-17.5cm}\caption{Enron long assets data: MCMC traces (left panels) and posterior densities (right panels) for $\mu$; with $\mu_1$ on the first row and
 $\mu_2$ on the second row.}\label{enron_fa_pars_2}
\end{center}
\end{figure}

\begin{figure}[htbp]
\begin{center}
\hspace{2cm}\includegraphics[width=16cm]{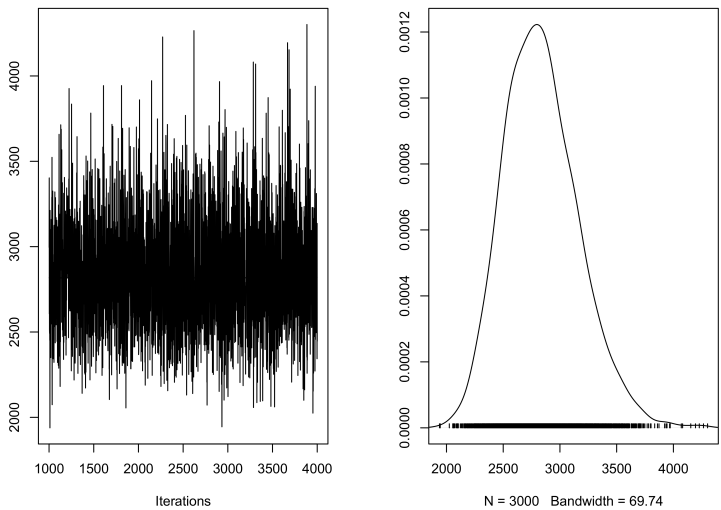}\\
\vspace{-13cm} \caption{Enron long term assets data: MCMC trace
(left panel) and posterior density (right panel) for
$\sigma^2$.}\label{enron_fa_pars_3}
\end{center}
\end{figure}

Focusing on the MCMC results, here we illustrate the outcomes of
the Enron long term assets data, since the results of the
remaining marginals are very similar to those presented. Figures
\ref{enron_fa_pars_1}, \ref{enron_fa_pars_2} and
\ref{enron_fa_pars_3} depict MCMC trace plots and posterior
densities, obtained using kernel density estimation from the R
package ``bayesmix" of \cite{Gr14}, for the parameters $\eta$,
$\mu$ and $\sigma^2$, respectively. We run the algorithm for 4,000
iterations, discarding the first 1,000 iterations as burn-in
period. The trace plots show that the chains are well mixing,
exploring freely the sample space and clearly reaching convergence
to the target distribution. Moreover, the unidentifiability
problem due to label switching, that may lead to biased estimates,
in our case does not occur. Finally, the posterior density plots
have regular forms and do not show multimodalities.

\subsection{PCC for Asset and Liability Data}

Following the procedure  described in Section \ref{CPE} we select
an appropriate pair copula decomposition for the D-vine. %As
%previously illustrated, in tree 1 the marginals are ordered
%according to the strongest pairwise dependencies (see Figure
%\ref{D_vine}), measured by Kendall's $\tau$.
For each one of the defaulted firms the order of
the marginals that maximizes the pairwise Kendall's $\tau$ indexes
in the first tree is $
    A_{C_{T}} \hspace{0.1cm} - \hspace{0.1cm} B_{C_{T}} \hspace{0.1cm} - \hspace{0.1cm} B_{L_{T}} \hspace{0.1cm} - \hspace{0.1cm} A_{L_{T}}.
    $

In Tables \ref{cirio_pcc_pars}, \ref{enron_pcc_pars},
\ref{parmalat_pcc_pars} and \ref{swissair_pcc_pars} we display,
for the defaulted firms, the list of pair
 copulas for each D-vine, the selected copula families, the copula parameters (one or two according to the type of copula) and the corresponding Kendall's
  $\tau$. The results are obtained using the R package ``CDVine''
  by
  \cite{BreSch13}.
From the selected copula families, we see evidence of different
types of asymmetric dependence. This demonstrates that the choice
of PCCs is appropriate, since it guarantees enough flexibility to
model the complex and asymmetric dependence structure of the data
at hand. Note that only the Cirio D-vine (Table
\ref{cirio_pcc_pars}) has none conditional independent variable
pairs. For these data the Genest and Favre (\cite{GenFav07})
independence test rejected independency for all the copulas
involved. An independence copula has been selected instead for
$c_{A_{C_{T}},B_{L_{T}}|B_{C_{T}}}$ in the second tree for
Parmalat and Swissair (Tables \ref{parmalat_pcc_pars} and
\ref{swissair_pcc_pars}), while
$c_{A_{L_{T}},B_{C_{T}}|B_{L_{T}}}$ has been identified as an
independence copula for Enron (Table \ref{enron_pcc_pars}). In
these cases the D-vine structure is simplified and we do not need
to estimate the parameters of the copula
$c_{A_{C_{T}},A_{L_{T}}|B_{C_{T}},B_{L_{T}}}$ in the third tree.
The presence of conditional independence in this last case
suggests a weak relationship between the current and long term
assets, given the values of liabilities. From the unconditional
pair copulas, we note an existing dependence between current and
long term assets or liabilities, and also a dependence between the
two different types of liabilities. A strong dependence in
conditional copulas instead may suggest imbalance, when current
assets are financed by long term liabilities, or a serious
liquidity problem, when long term assets are financed by current
liabilities. These situations need particular attention, because
they may prelude to the default of the firm.

For the Sysco company the order in the first tree is \linebreak
$A_{C_{T}} \hspace{0.1cm} - \hspace{0.1cm} A_{L_{T}}
\hspace{0.1cm} - \hspace{0.1cm} B_{L_{T}} \hspace{0.1cm}
-\hspace{0.1cm} B_{C_{T}}.$ In Table \ref{sysco_pcc_pars} we
display the list of pair
 copulas for the D-vine, the selected copula families, the copula parameters (one or two according to the type of copula) and the corresponding Kendall's $\tau$.
In this case, an independence copula has been selected for
$c_{A_{C_{T}},B_{L_{T}}|A_{L_{T}}}$. The D-vine structure is
simplified and we do not need to estimate the parameters of the
copula $c_{A_{C_{T}},B_{C_{T}}|A_{L_{T}},B_{L_{T}}}$ in the third
tree. The presence of conditional independence in this last case
suggests a weak relationship between the current asset and
liabilities, given the values of the long term ones.

%Figure \ref{enron_pcc_tree} shows the D-vine tree plot for the
%Enron data and contains the trees of the D-vine. It is obtained
%with the R package CDVine by \cite{BreSch13}. The squares
%represent nodes that are variables, while the lines represent arcs
%that are dependencies between variables. The names of the nodes
%may be read in the  squares, and the pair copula families and
%Kendall's $\tau$ values corresponding to pair copula parameters
%can be read in the edge labels. The thicker the line the higher
%the dependence between the variables represented by the nodes.
%Only the Cirio D-vine contains all three trees, while the D-vines
%of the remaining data contain two trees only, because of
%simplification derived by conditional independence.

\begin{table}[htbp]
 \caption{Cirio: selected copulas and D-vine PCC parameters. SBB1, BB7 and BB8 are, respectively, the Survival Clayton-Gumbel,
 the Joe-Clayton and the Joe-Frank copulas, that are Ar\-chi\-me\-dean copula families with two parameters.}\label{cirio_pcc_pars}
  \vspace{0.3cm}
  \centering
    \begin{tabular}{ccccc}
   \hline
    \multicolumn{5}{c}{\textbf{Cirio: Pair Copula Parameters of the D-Vine}}  \\
  \hline
    \textit{Copulas} & \textit{family} & \textit{parameter 1} & \textit{parameter 2} & \textit{Kendall's $\tau$} \\
    \hline
    $c_{A_{C_{T}},B_{C_{T}}}$   & SBB1   & 0.0010 & 3.3814 & 0.7044 \\
    $c_{B_{C_{T}},B_{L_{T}}}$   & BB8   & 1.2579 & 0.9902 & 0.1186 \\
    $c_{B_{L_{T}},A_{L_{T}}}$   & BB7   & 1.1195 & 4.7016 & 0.6985 \\
    $c_{A_{C_{T}},B_{L_{T}}|B_{C_{T}}}$ & Frank & 7.2222 & N/A & 0.5718 \\
    $c_{A_{L_{T}},B_{C_{T}}|B_{L_{T}}}$ & Normal & -0.0337 & N/A & -0.0214 \\
    $c_{A_{C_{T}},A_{L_{T}}|B_{C_{T}},B_{L_{T}}}$ & Frank & -8.9557 & N/A & -0.6353 \\
    \hline
    \end{tabular}
   \end{table}

\begin{table}[htbp]
\caption{Enron: selected copulas and D-vine PCC parameters. SBB8
and BB8 are, respectively, the Survival Joe-Frank and the
Joe-Frank copulas,
     Archimedean copula families with two parameters.}\label{enron_pcc_pars}
      \vspace{0.3cm}
  \centering
    \begin{tabular}{ccccc}
   \hline
    \multicolumn{5}{c}{\textbf{Enron: Pair Copula Parameters of the D-Vine}}  \\
  \hline
    \textit{Copulas} & \textit{family} & \textit{parameter 1} & \textit{parameter 2} & \textit{Kendall's $\tau$} \\
    \hline
    $c_{A_{C_{T}},B_{C_{T}}}$   & Student's t & 0.9868 & 7.6539 & 0.8963 \\
    $c_{B_{C_{T}},B_{L_{T}}}$   & SBB8  & 6.0000 & 0.3924 & 0.2761 \\
    $c_{B_{L_{T}},A_{L_{T}}}$   & BB8   & 5.9831 & 0.9979 & 0.7208 \\
    $c_{A_{C_{T}},B_{L_{T}}|B_{C_{T}}}$ & Rotated Clayton & -1.4730 & N/A & -0.4241 \\
    $c_{A_{L_{T}},B_{C_{T}}|B_{L_{T}}}$ & Independence & N/A     & N/A & 0 \\
    \hline
    \end{tabular}
    \end{table}

\begin{table}[htbp]
 \caption{Parmalat: selected copulas and D-vine PCC parameters. BB1 is  the Clayton-Gumbel copula,
     Archimedean copula family with two parameters.}\label{parmalat_pcc_pars}
      \vspace{0.3cm}
  \centering
    \begin{tabular}{ccccc}
   \hline
    \multicolumn{5}{c}{\textbf{Parmalat: Pair Copula Parameters of the D-Vine}}  \\
  \hline
    \textit{Copulas} & \textit{family} & \textit{parameter 1} & \textit{parameter 2} & \textit{Kendall's $\tau$} \\
    \hline
    $c_{A_{C_{T}},B_{C_{T}}}$   & BB1   & 0.4325 & 4.2015 & 0.8043 \\
    $c_{B_{C_{T}},B_{L_{T}}}$   & Normal & 0.9998 & N/A & 0.9898 \\
    $c_{B_{L_{T}},A_{L_{T}}}$   & Clayton & 1.2256 & N/A & 0.3800 \\
    $c_{A_{C_{T}},B_{L_{T}}|B_{C_{T}}}$ & Independence & N/A     & N/A & 0 \\
    $c_{A_{L_{T}},B_{C_{T}}|B_{L_{T}}}$ & Frank & -7.3657 & N/A & -0.5778 \\
    \hline
    \end{tabular}
   \end{table}

\begin{table}[htbp]
  \caption{Swissair: selected copulas and D-vine PCC parameters. BB7 and SBB8 are, respectively, the Joe-Clayton and the Survival Joe-Frank copula,
   Archimedean copula families with two parameters}\label{swissair_pcc_pars}
  \vspace{0.3cm}
  \centering
    \begin{tabular}{ccccc}
   \hline
    \multicolumn{5}{c}{\textbf{Swissair: Pair Copula Parameters of the D-Vine}}  \\
  \hline
    \textit{Copulas} & \textit{family} & \textit{parameter 1} & \textit{parameter 2} & \textit{Kendall's $\tau$} \\
    \hline
    $c_{A_{C_{T}},B_{C_{T}}}$   & BB7   & 2.4309 & 5.3880 & 0.7267 \\
    $c_{B_{C_{T}},B_{L_{T}}}$   & SBB8  & 1.0081 & 1.0000 & 0.0047 \\
    $c_{B_{L_{T}},A_{L_{T}}}$   & BB7   & 1.0010 & 2.9494 & 0.5959 \\
    $c_{A_{C_{T}},B_{L_{T}}|B_{C_{T}}}$ & Independence & N/A     & N/A & 0 \\
    $c_{A_{L_{T}},B_{C_{T}}|B_{L_{T}}}$ & Rotated Joe & -2.3405 & N/A & -0.4222 \\
    \hline
    \end{tabular}
  \end{table}

\begin{table}[htbp]
  \caption{Sysco: selected copulas and D-vine PCC parameters. BB1 and  BB6 are, respectively, the Clayton-Gumbel and the Survival Joe-Gumbel copulas,
  that are  Archimedean copula family with two parameters.}\label{sysco_pcc_pars}
  \vspace{0.3cm}
  \centering
    \begin{tabular}{ccccc}
   \hline
    \multicolumn{5}{c}{\textbf{Sysco: Pair Copula Parameters of the D-Vine: }}  \\
  \hline
    \textit{Copulas} & \textit{family} & \textit{parameter 1} & \textit{parameter 2} & \textit{Kendall's $\tau$} \\
    \hline
    $c_{A_{C_{T}},A_{L_{T}}}$   &  BB6    & 6 & 1.9387 & 0.8569 \\
    $c_{A_{L_{T}},B_{L_{T}}}$   &  Frank  & 298.314
  &  N/A & 0.9867 \\
    $c_{B_{L_{T}},B_{C_{T}}}$   & BB1  & 4.4714   & 1.7693 & 0.8253 \\
    $c_{A_{C_{T}},B_{L_{T}}|A_{L_{T}}}$ & Independence & N/A     & N/A & 0 \\
    $c_{A_{L_{T}},B_{C_{T}}|B_{L_{T}}}$ & Normal &-0.0270  & N/A & -0.0172 \\
    \hline
    \end{tabular}
  \end{table}

%\begin{figure}
%\begin{center}
%  \includegraphics[width=13cm]{enron_vine_tree_plot_1.pdf}\\
%   \includegraphics[width=13cm]{enron_vine_tree_plot_2.pdf}\\
%  \end{center}
%  \vspace{-2cm}\caption{Enron data: D-vine PCC tree.}\label{enron_pcc_tree}
%\end{figure}

\subsection{Probability of Default Estimation}

\begin{figure}
\begin{center}
  \includegraphics[width=22cm]{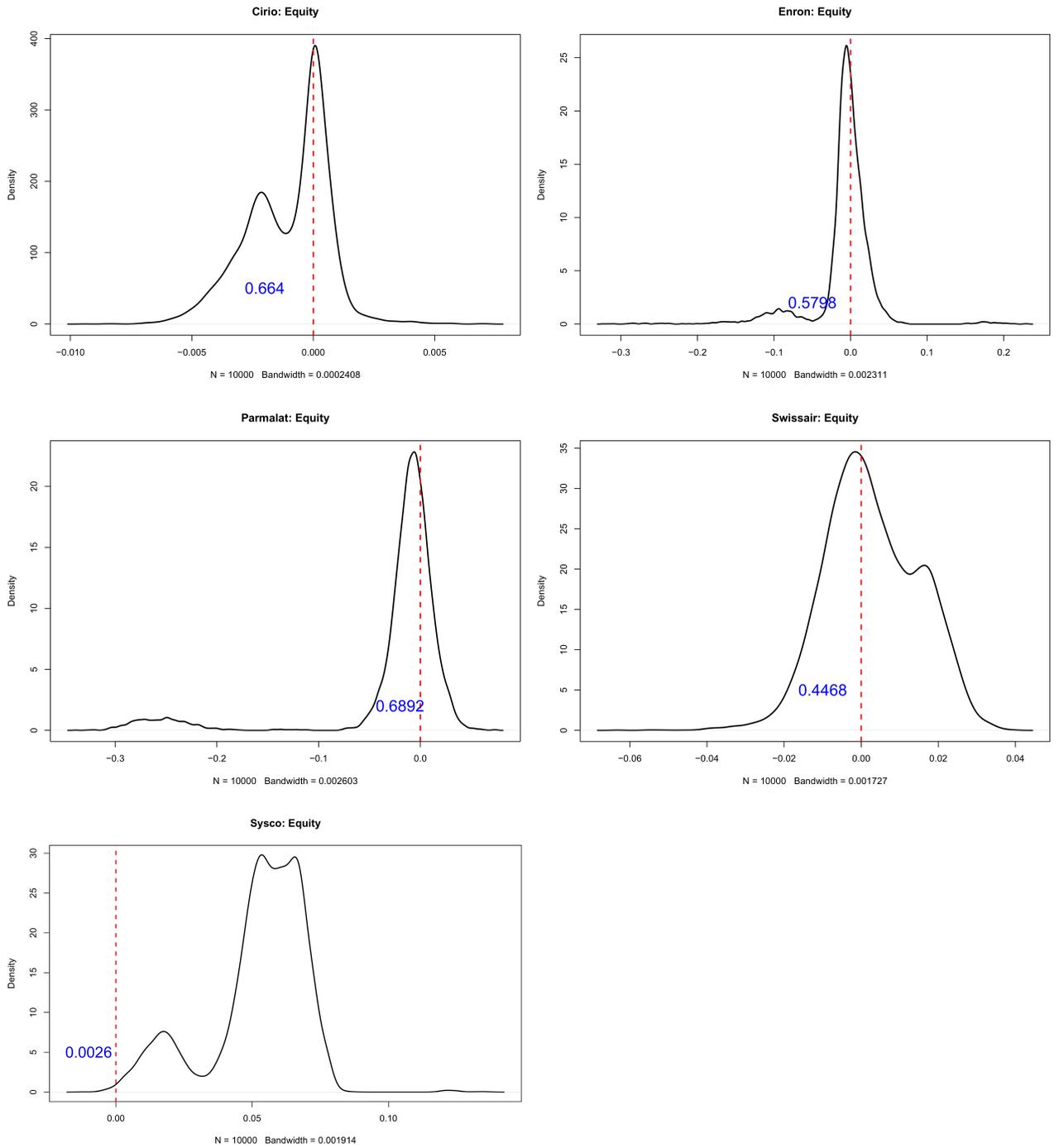}
  \vspace{-10cm}\caption{Equity densities of Cirio (top left), Enron (top right),
  Parmalat (middle left), Swissair (middle right) and Sysco (bottom). The value of the PD is written on the corresponding densities.}\label{equity_PD}
  \end{center}
\end{figure}

To estimate the PD we follow the methodology described in  Section
\ref{mod1}. For each firm we generate 10,000 simulations from the selected D-vine to
obtain the equity distribution and the PD.
Figure \ref{equity_PD} depicts the equity densities of Cirio, Enron, Parmalat, Swissair and Sysco, respectively on the top left, top right, middle left, middle right and bottom panel. The Figure was obtained using kernel density estimation.
The value of the PD is
written in the relevant plot and corresponds to the area
under the curve where the e\-qui\-ty is zero or negative. The PD
values are very high for all defaulted firms, lying in the range
$0.4468-0.6892$; in contrast, the Sysco PD is only $0.0026$, as we
expected for a healthy company, where the probability of going
bankrupt is very low. We notice that the PD values of Enron and
Swissair are slightly lower than the other defaulted firms.
However, the available balance sheet data did not include the last
year of activity of Enron and Swissair. This might have affected
the final results, since the inclusion of the last year's data
would certainly have increased the corresponding PD values.

For comparative purposes we contrasted our results with those
obtained applying the original Z-score proposed by \cite{Altman68}, e.g. to the Enron company.
\begin{figure}
\begin{center}
  \includegraphics[width=8cm]{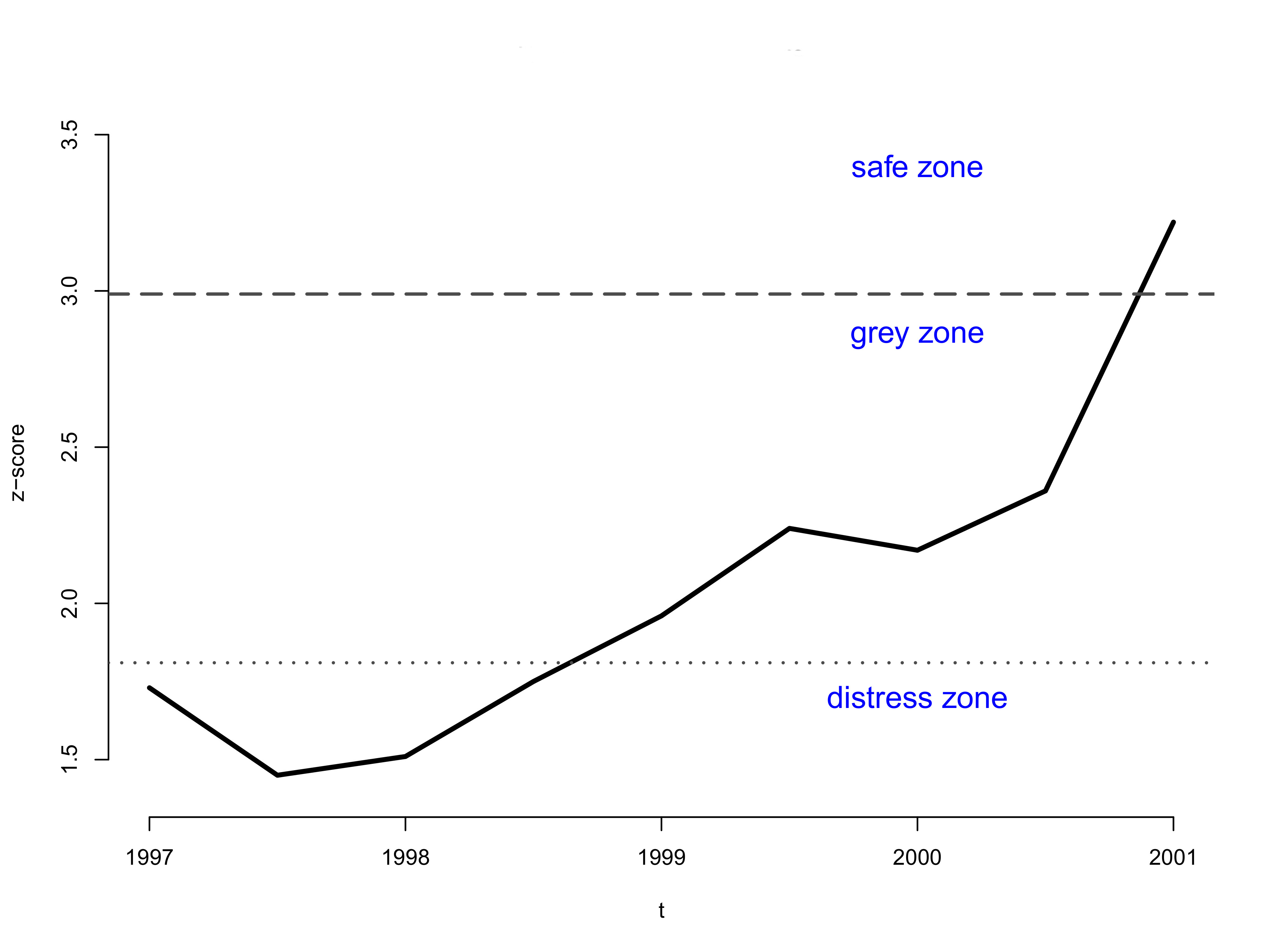}
  \caption{Altman's Z-score for Enron data}\label{Enron_Altman_z}
  \end{center}
\end{figure}
Figure \ref{Enron_Altman_z} shows the line plot of Altman's
Z-score for the time horizon between 1997 and 2001. According to
\cite{Altman68}, a company is considered to be in the ``safe''
zone (healthy) when $z > 2.99$, it is in the ``grey'' zone
(moderate risk of default) when $1.81 < z < 2.99$, and it is in
the ``distress'' zone (high danger of default) when $z < 1.81$.
Altman's Z-score is clearly not able to predict the failure of the
Enron company, since it locates the firm in the distress zone only
until 1998; subsequently, from 1999 to 2000, it moves the firm to
the grey zone (erroneously suggesting an improved performance);
and finally (when the actual
 default actually occurred) places the firm in the safe area, with a \emph{z}-score of 3.22.
Besides, in the considered period Altman's Z-score is not
decreasing, but even rising, leading to completely misleading
conclusions.

The Z-score's inability of predicting the default is due to the
fact that, unlike the PCC model, it does not consider the
dependence pattern among the different components of the balance
sheet data. On the contrary, the proposed PCC model allows us to
measure the dependencies and to detect in advance alarming
situations, which are not identifiable using other traditional
models. Moreover, our PCC approach permits to adopt different and
more suitable marginal distributions, better reflecting the
structure of the data at hand.

Moreover, the calculation of Altman's Z-score, unlike the PCC
model, involves balance sheet data as well as economic and income
data, without analysing the relationship among these quantities.
For this reason, the Z-score might mask dangerous default risks,
classifying a company as ``safe'', when it is truly in distress.

\section{Summary and Conclusions}

The aim of this paper was to propose a novel methodology for PD
evaluation. Our final goal was to calculate the PD of large firms
using their balance sheet data. We measured the firm value via a
contingent claim, whose pricing function may be expressed using
copulas. The marginals are given by the current and long term
assets and  liabilities. Hence, the equity function is expressed
by a 4-dimensional D-vine copula. To test the performance of our
methodology we applied it to four fraudulent defaulted stocks and
to the data of a healthy firm. In order to estimate the marginals
we employed a Bayesian mixture model, able to model the presence
of two clusters in the asset as well as liability data. The
structure of the marginals in defaulted firms reflects the choices
of the management, trying to balance high and low accounting items
during the period before the default. Considering the copula, we
chose to employ PCCs, because they allow for a great flexibility
in modelling the dependence structure of the marginals. As
demonstrated by the results, the pair copulas selected in the
D-vines belong to different families and describe various types of
dependence. The analysis of these dependencies in defaulted firms
data already reveals substandard loans and situations of serious
imbalance due to liquidity issues, especially when the firm tries
to balance long term assets with current liabilities. Finally, we
calculated the PD of the five considered firms, simulating from
the D-vines and obtaining the e\-qui\-ty distributions. The final
results show a high PD for the defaulted firms, suggesting their
forthcoming bankruptcy. The PD of the Sysco company is instead
much lower than those of the defaulted firms, denoting a good
performance. A traditional indicator like Altman's Z-score may be
incapable of predicting the risk of default, since it is not
flexible enough and it does not incorporate the dependence
structure of the involved quantities. On the contrary, the
proposed methodology has proven to be successful in the evaluation
of PD and would certainly benefit analysts and managers, advising
them to take actions against a potential bankruptcy.

Possible extensions of our work include the estimation of the
whole model in a full Bayesian
framework, the application of
nonparametric approaches for PCCs,
 the use of balance indicators
instead of accounting items and the use of a our methodology to
analyze the contagion in sectors of activity. Finally, it would be
interesting to apply our approach to Altman's Z-score, to model
the dependence between the different quantities involved.

\section*{Acknowledgements}

The work of the second author was partially supported by MIUR,
Italy, PRIN MISURA  2010RHAHPL.  The authors are thankful to
Michela Magliacani and Dennis Montagna for helpful comments and
suggestions. We are grateful to the Editor and the two anonymous
referees for valuable comments.

%\section*{References}

\end{document}